\documentclass[letterpaper,twocolumn,10pt]{article}
\usepackage{usenix-2020-09}
\usepackage{minorrev}

\usepackage{tikz}
\usepackage{soul}
\usepackage{amsmath}

\usepackage{hhline}
\usepackage{multirow}
\usepackage{colortbl}

\begin{document}

\date{}

\title{``I wasn’t sure if this is indeed a security risk'': Data-driven Understanding\\ of Security Issue Reporting in GitHub Repositories of Open Source npm Packages}

\author{
{\rm Rajdeep Ghosh}\\
IIT Kharagpur \\
ghoshrajdeep2000@gmail.com
\and
{\rm Shiladitya De}\\
IIT Kharagpur \\
shiladityade.bwn2001@gmail.com
\and
{\rm Mainack Mondal}\\
IIT Kharagpur \\
mainack@cse.iitkgp.ac.in
} 

\maketitle

\begin{abstract}

The npm (Node Package Manager) ecosystem is the most important package manager for JavaScript development with millions of users. Consequently, a plethora of earlier work investigated how vulnerability reporting, patch propagation, and in general detection as well as resolution of security issues in such ecosystems can be facilitated. \newtext{However, understanding the ground reality of security-related issue reporting by users (and bots) in npm--along with the associated challenges--has been relatively less explored at scale.}

In this work, we bridge this gap by collecting 10,907,467 issues reported across GitHub repositories of 45,466 diverse npm packages. We found that the tags associated with these issues indicate the existence of only 0.13\% security-related issues. However, our approach of manual analysis followed by developing high-accuracy machine learning models identify 1,617,738 security-related issues which are not tagged as security-related (14.8\% of all issues) as well as 4,461,934 comments made on these issues. We found that the bots which are in wide use today might not be sufficient for either detecting or offering assistance with these issues. Furthermore, our analysis of user-developer interaction data hints that many user-reported security issues might not be addressed by developers---they are not tagged as security-related issues and might be closed without valid justification. Consequently, a correlation analysis hints that the developers quickly handle security issues with known solutions (e.g., corresponding to CVE, or with a suggested solution). However, security issues without such known solutions \newtext{(even with reproducible code)}
might not be resolved, hinting at a need for better-automated assistance for npm developers to address security issues. Our findings offer actionable insights for improving security management in open-source ecosystems, highlighting the need for smarter tools and better collaboration. The data and code for this work is available at \url{https://doi.org/10.5281/zenodo.15614029}

\end{abstract}

\section{Introduction}
Node Package Manager or \textit{npm}~\cite{npm} is a widely used ecosystem to manage and distribute JavaScript software packages (i.e., libraries). Serving as a central hub for open-source development, it powers diverse applications and services. The npm ecosystem is defined by its dense interdependencies, where packages build upon others, forming a highly interconnected network. However, with increasing interdependency, the risk of propagation of security vulnerabilities increases. Nevertheless, as an open-source platform, npm benefits from a collaborative community, which plays a crucial role in identifying, addressing, and mitigating these vulnerabilities.

Previous research has extensively investigated security vulnerability management systems in prominent package ecosystems such as npm, PyPI, and RubyGems. They focused on vulnerability reporting, propagation, and resolution ~\cite{Decan2019,10172868,Zerouali2022,10.1145/3196398.3196401,10.1007/s10664-021-09959-3}. Upstream vulnerabilities, patch delivery delays, and the significance of regular dependency updates all represent considerable risks, according to research ~\cite{10.1145/3597503.3639230, 10172868}. Furthermore, the prior work investigated the function of bots in automating processes, increasing productivity, and supporting open-source software (OSS) workflows, with varied results regarding their usefulness and limitations in security management ~\cite{10.5555/3155562,10.5555/3155562.3155577,10.1145/3274451}.

However, majority of these previous works focus on proposing the \textit{automated} security vulnerability management tools and techniques. They often overlook what techniques are \textit{actually} used in real world. Prior work also did not shed light on whether public (i.e., open source) aspect of ecosystems like npm actually helps in detecting security flaws and mitigating them. Specifically, it is not clear whether users or bots even report/suggest mitigation of security issues in these ecosystems and if those reports have any impact on improving the security of the npm packages. \newtext{Answering these quite important unanswered questions will help the npm community (both developers and users) evaluate the effectiveness of currently deployed mechanisms for reporting and addressing security issues and identify areas for improvement.} 

In this work, for the first time, we answer these questions via large-scale user-generated data collection and analysis. We focus on publicly available npm packages and their GitHub repositories (since GitHub provides security issues and comments for these npm packages). \snewtext{By default, issues (both  security and non-security), on public GitHub repositories are publicly visible~\cite{10.1145/2811257, uie_users_change_settings}. While features like Private Vulnerability Reporting (PVR) enable private disclosure, they are relatively recent and are not enabled by default~\cite{githubPVR}. Thus, it is not yet widely adopted.} 

We systematically chose 45,466 npm packages with different numbers of \textit{dependents} \newtext{(i.e., how many other packages they are used by}
---a measure of popularity and impact). For these packages, we collected 37,278 GitHub repositories, and from these repositories, we collected 10,907,467 issues raised in GitHub. Interestingly, only 0.13\% of these issues were tagged by the GitHub repository owners as security-related. \newtext{However, by leveraging machine learning, we found a significant 14.8\% of user-reported security issues which are not tagged with any security-related tags.}
Overall, we collected and analyzed 1,617,738 security-related issues (as well as 4,461,934 comments on those issues) for these repositories. Furthermore, we leverage this data to analyze (i) how each security-related issue is handled during the creation, discussion, and resolution phase (ii) what are the functionalities of bots which are used today in GitHub repositories of npm packages. Overall, we took a deep dive into this large-scale data of security-related issues in npm packages---our in-depth mixed-method analysis (complemented by creating novel machine learning-based detection models) of these issues uncovered a hierarchy of themes for interaction between users/developers while handling these issues as well as factors that affected the resolution of such issues. Specifically, in this work, we answer four key research questions.

\textit{\textbf{RQ1 }How prevalent is reporting security-related issues in GitHub repositories of npm packages?} 

\textit{\textbf{RQ2 }How effective are bots in detecting and addressing security-related issues?} 

\textit{\textbf{RQ3 }How do the users interact with the GitHub repository maintainers for npm packages while reporting security-related issues?}

\textit{\textbf{RQ4 } What factors correlate with the resolution of security-related issues?}

\newtext{While investigating these questions, we found that across npm packages with varying numbers of dependents, around 10\%–15\% of issues are security-related.} In fact, more than 23\% of the issues did not receive any comments (Section~\ref{sec:rq1}). Interestingly, although bots in these repositories help to report security issues, their effectiveness is rather limited, as hinted by the fact that less than 0.1\% security issues are tagged by bots in our data. 
In fact, our in-depth systematic analysis \newtext{reveals} that there is a general lack of security-focused bot usage which leverage techniques like static analysis or machine learning (Section~\ref{sec:rq2}).  In fact, the number of such bots are also quite limited in our dataset. Moreover, this concerning situation is also present within user-developer interactions. The user-reported security-related issues are often ignored by developers and can be closed without a valid justification (since it becomes stale) (Section~\ref{sec:rq3}). Our correlation analysis identifies a potential reason---only when the user reported issues contain a potential fix (via pull request) or include a CVE \newtext{---publicly accepted set of already reported security vulnerabilities (CVE) and weaknesses (CWE), the npm developers resolve this issue (unless otherwise specified, we treat CVE and CWE references equivalently throughout the paper). Otherwise, even} when reproducible code for a security-issue is included in a reported issue it is more likely to become stale without any resolution. Our findings hint at the need for improving bots as well as resolution techniques for reported security issues in repositories of npm packages. 

\vspace{1mm}
\noindent
\textbf{Limitations:} \newtext{Due to computational constraints and API rate limits, collecting GitHub links for millions of npm packages was challenging. As a trade-off, we followed a practical and coverage-driven approach using dependent-based stratified sampling. Since the number of dependents indicates the potential impact of security issues, we included all high-impact packages (> 100 dependents) and large samples of low-impact packages (Table-~\ref{npm_packages_stats_1}). Thus, although we might have missed some packages, security implications for those missing packages might be limited.} Our approach has the potential to miss security issues which are not reported to GitHub. However, we already identified a significant number of reported security issues (which are not tagged with security-related tags), and this shortcoming can only increase the number of such issues. Thus, our results are potentially lower bound on actual security-related issues. \snewtext{One possible limitation is that some context-specific tags may not be captured by our Word2Vec based methodology. However, our final machine learning model to detect security issues was trained (and used) on the full issue descriptions (both title and body). Thus, we ensure that security-related issues are robustly classified, even when tags are missing or applied inconsistently.}
Furthermore, some inaccuracy is associated with machine learning models used to identify potential security-related issues and related themes. We attempted to reduce this concern by hyperparameter adjustment, rigorous model selection, and manual validation in line with best practices. Furthermore, our thematic saturation attained during thematic analysis suggests that there will be a very small number of such unidentified themes after the application of our machine learning models over millions of issues and comments, if any, which may have a minor impact on the validity of the presented results. In summary, we uncovered interesting and potentially generalizable factors affecting the resolution of security-related issues using our ecologically valid dataset despite these limitations. 

\section{Related Work} \label{sec:rel_work}
\noindent We review the related work along four broad dimensions: security vulnerability management in package ecosystems, analysis of bots in developer workflows, \snewtext{interaction with developers/maintainers} and detection of malicious packages in repositories. 

\vspace{1mm}

\noindent
\textbf{Security vulnerability management in package ecosystems}: Previous work studied popular package ecosystems, e.g., npm, PyPI, RubyGems. Many of these studies have focused on how security vulnerabilities in one package \newtext{make} its dependent packages vulnerable~\cite{Decan2019,10172868,Zerouali2022,10.1145/3196398.3196401,10.1007/s10664-021-09959-3}. 
Alfadel et al.~\cite{10.1145/3571848} revealed that many Node.js applications rely on packages with undisclosed vulnerabilities, emphasizing the need for faster remediation. Similarly, findings from other works on npm, RubyGems and Golang ecosystems stress getting timely updates and patches about vulnerabilities in the dependencies of a package~\cite{Zerouali2022,10172868,10.1145/3597503.3639230}. \newtext{The study by B\"uhlmann et al.~\cite{10.1145/3477314.3507123} examines developer responses to security issues in Java repositories,} while our study, in contrast, collected and analyzed security concerns raised by the users themselves via issues reported in GitHub repositories of open source npm packages \newtext{often involving high-accuracy ML models }. These community-reported security concerns from the real world, unlike previous work, are often explicitly not connected to the known vulnerabilities detected in the dependencies (e.g., via CVE). 

\vspace{1mm}
\noindent
\textbf{Analyzing bots in developer workflows:} Bots (programs to automate tasks) have been extensively studied for their ability to automate tasks, enhance productivity, and support decision-making in various domains, including open-source software (OSS) ~\cite{10.1145/2950290.2983989,8239928}. On platforms like GitHub, bots are frequently employed for tasks such as continuous integration, dependency management, and collaborative modeling~\cite{10.5555/3155562,10.1145/3183519,10.5555/3155562.3155577}. Prior studies have investigated their impact on developer workflows, revealing alterations in commit activity and pull request closure times following bot implementation. Wessel et al. found that out of 351 GitHub projects, 26\% utilised bots and a recent follow-up study even introduced bots to address negative impacts on contributions ~\cite{10.1145/3274451,10.1145/3387940.3391504}. However, none of these works considered either the diverse functions that bots play in a real-world ecosystem like GitHub repositories of npm packages or the roles bots played in addressing security issues within OSS (specifically, npm). For the first time, this work bridges this gap and investigates the effectiveness of these bots in identifying and addressing security vulnerabilities. To do so, we built on another line of research---bot detection. Methods like BotHunter and BIMAN leveraged machine learning to identify bots in GitHub, whereas Golzadeh et al. developed a ground-truth dataset of GitHub issues and PR comments, to detect bot accounts~\cite{10.1145/3524842.3527959, 10.1145/3379597.3387478, GOLZADEH2021110911}. We used these techniques to detect bots within issues and comments posted in the npm package repositories on GitHub and uncover that the functionalities of existing bots are not enough to address security issues within these repositories. 

\vspace{1mm}

\noindent \snewtext{\textbf{Interaction with Developer/maintainer in package repositories:} A significant amount of prior works investigated interaction between developers to understand collaboration patterns~\cite{10.1007/s11219-022-09598-x}, on boarding mechanisms~\cite{10.1145/3368089.3409746}, sentiment difference~\cite{10.1145/3194932.3194936} and community dynamics~\cite{10.1145/3449093}. However, the focus of these works typically has been on general developer collaboration and project management practices. Rather, our results are more specific to security, e.g., impact of bots and CVEs.}

\vspace{1mm}

\noindent \textbf{Detection of malicious packages in package repositories} 
: A plethora of previous work aimed to detect malicious packages in popular package repositories such as npm and PyPI. They often use machine learning models, graph analysis, or static and dynamic analysis techniques~\cite{10.1145/3627106.3627138,10.1145/3705304,article_vulnet,10.1145/3510003.3510104,10.1145/3691620.3695492, 10.1145/3510003.3510142,10.1145/3650212.3680397}. Out of them AMALFI by Sejfia and Schafer is specifically tailored to detect malicious packages in JavaScript and TypeScript ecosystems~\cite{10.1145/3510003.3510104}. Furthermore, VulNet ~\cite{article_vulnet}, PatchFinder ~\cite{10.1145/3650212.3680305}, Holmes ~\cite{10.1145/3597503.3639582}, and Ranger ~\cite{10298471} have introduced innovative solutions to enhance vulnerability prioritization, patch tracing, and secure version restoration for ecosystem-wide security management. Going one step further, Ferreira et al.~\cite{9402108} introduced a lightweight, permission-based system for Node.js applications, making it significantly more challenging to exploit malicious packages. These prior works generally focused on finding (and sometimes mitigating) security vulnerabilities by analyzing codebases. On the contrary, our work focuses on the security concern-related community feedback received on these packages, presumably about the security issues identified by users, making our work complementary to this prior body of work on automated malicious package detection. 

Next we will present our approach of collecting community-reported security issues at scale from a stratified sample of tens of thousands of npm packages.

\section{Collecting Data on Issues Reported in GitHub Repositories of npm packages}\label{sec:data_creation}

\noindent For our study, we needed to gather large-scale data about security issues reported for real-world GitHub repositories of npm packages. In this section, we detail our approach. 

\subsection{Selecting npm packages}

\noindent \textbf{Collecting package data from npm:} npm (Node Package Manager)~\cite{npm} is the largest and most popular repository for \newtext{JavaScript} libraries (called packages) today. We downloaded the list of the entire 4.3 million public \newtext{JavaScript} packages hosted on npm during May 2024. 
Furthermore, for each package, we also downloaded the \textit{dependents}---packages which depend on this package using the API~\cite{registry}. However, collecting and analyzing issues from all 4.3 million packages was computationally difficult. To that end, we decided to create a stratified sample of these packages based on the number of dependents (i.e., how many other packages used these packages). The reason is simple---intuitively, the importance of reported security related issues for a package is correlated with the number of dependents (used in earlier work~\cite{10172868}). 

\begin{table}[]
\scriptsize
\centering
\begin{tabular}{|lll|l|}
\hline
\rowcolor[HTML]{C0C0C0} 
\multicolumn{1}{|l|}{\cellcolor[HTML]{C0C0C0}\textbf{\begin{tabular}[c]{@{}l@{}}\# dependent \\ packages\end{tabular}}} & \multicolumn{1}{l|}{\cellcolor[HTML]{C0C0C0}\textbf{\begin{tabular}[c]{@{}l@{}}\# Packages \\ available\end{tabular}}} & \textbf{\begin{tabular}[c]{@{}l@{}}\# Packages \\ sampled\end{tabular}} & \textbf{\begin{tabular}[c]{@{}l@{}}\# Packages with\\ GitHub link\end{tabular}} \\ \hline
\multicolumn{1}{|l|}{0}                                                                                                & \multicolumn{1}{l|}{3,457,606}                                                                                          & 20,000                                                                   & 7023                                                                                                         \\ \hline
\multicolumn{1}{|l|}{1-10}                                                                                             & \multicolumn{1}{l|}{462,379}                                                                                           & 20,000                                                                   & 13,766                                                                                                        \\ \hline
\multicolumn{1}{|l|}{10-100}                                                                                           & \multicolumn{1}{l|}{41,982}                                                                                            & 20,000                                                                   & 15,897                                                                                                        \\ \hline
\multicolumn{1}{|l|}{100-500}                                                                                          & \multicolumn{1}{l|}{7,899}                                                                                             & 7,899                                                                    & 6,557                                                                                                         \\ \hline
\multicolumn{1}{|l|}{500-1,000}                                                                                         & \multicolumn{1}{l|}{1,132}                                                                                             & 1,132                                                                    & 1,117                                                                                                         \\ \hline
\multicolumn{1}{|l|}{> 1,000}                                                                                           & \multicolumn{1}{l|}{1,112}                                                                                             & 1,112                                                                    & 1,106                                                                                                         \\ \hline
\multicolumn{2}{|r}{\textbf{Total}} & \multicolumn{1}{|l|}{70,143} & 45,466 \\ \hline

\end{tabular}
\caption{We bucketed the npm packages by number of dependents. We present the number of total as well as sampled packages from each bucket alongwith number of packages where GitHub link was available.}
\label{npm_packages_stats_1}
\end{table}

\vspace{1mm}
\noindent \textbf{Stratified sampling of npm packages:} We simply divided the 4.3 million npm packages \newtext{into six buckets}
based on their number of dependents: 0, 1–10, 10–100, 100–500, 500–1,000, and greater than 1,000. The number of repositories for each bucket is reported in Table~\ref{npm_packages_stats_1}. We note that buckets with a lower number of dependents contained an overwhelming majority of the packages, making it impractical to include them all in our analysis. Thus, we simply randomly sampled 20,000 packages from each bucket (if a bucket contained less than 20,000 packages, we included all). In the end, in our sample, we ended up considering all the packages with more than 100 dependents and 60,000 packages with less than 100 dependents (Table~\ref{npm_packages_stats_1})---in total we curated 70,143 packages. Few examples of packages from each bucket is given in Table~\ref{pop_bucket} of Appendix~\ref{app_sec2}.

\subsection{Collecting issue data from GitHub for selected npm packages}

\begin{table}[]
\scriptsize
\centering
\begin{tabular}{|l|l|l|l|}
\hline
\rowcolor[HTML]{C0C0C0} 
{\color[HTML]{000000} \textbf{\begin{tabular}[c]{@{}l@{}}\# dependent \\ packages\end{tabular}}} & {\color[HTML]{000000} \textbf{\textbf{\begin{tabular}[c]{@{}l@{}}\# Sampled Packages \\with GitHub link\end{tabular}}}} & {\color[HTML]{000000} \textbf{\begin{tabular}[c]{@{}l@{}}\# Unique GitHub \\ repositories\end{tabular}}} & {\color[HTML]{000000} \textbf{\begin{tabular}[c]{@{}l@{}}\# Total \\ issues\end{tabular}}}  \\ \hline
{\color[HTML]{000000} 0}                                                                        & {\color[HTML]{000000} 7,023}                & {\color[HTML]{000000} 6,755}                                                                             & {\color[HTML]{000000} 1,323,970}                                                                                                                   \\ \hline
{\color[HTML]{000000} 1-10}                                                                     & {\color[HTML]{000000} 13,766}               & {\color[HTML]{000000} 12,186}                                                                            & {\color[HTML]{000000} 2,908,173}                                                                                                                \\ \hline
{\color[HTML]{000000} 10-100}                                                                   & {\color[HTML]{000000} 15,897}               & {\color[HTML]{000000} 11,687}                                                                            & {\color[HTML]{000000} 3,687,408}                                                                                                          \\ \hline
{\color[HTML]{000000} 100-500}                                                                  & {\color[HTML]{000000} 6,557}                & {\color[HTML]{000000} 5,015}                                                                             & {\color[HTML]{000000} 1,684,658}                                                                                                        \\ \hline
{\color[HTML]{000000} 500-1000}                                                                 & {\color[HTML]{000000} 1,117}                & {\color[HTML]{000000} 832}                                                                              & {\color[HTML]{000000} 344,343}                                                                                                          \\ \hline
{\color[HTML]{000000} > 1000}                                                                   & {\color[HTML]{000000} 1,106}                & {\color[HTML]{000000} 803}                                                                              & {\color[HTML]{000000} 958,915}                                                                                                                        \\ \hline
\multicolumn{1}{|l|}{{\color[HTML]{000000} \textbf{Total}}}                                              & {\color[HTML]{000000} \textbf{45,466}}               & {\color[HTML]{000000} \textbf{37,278}}                                                                            & {\color[HTML]{000000} \textbf{10,907,467}}                                                                                                    \\ \hline
\end{tabular}
\caption{Number of issues collected from GitHub pages of our sample set of npm packages.}
\label{stats_data_issues_collected_1}
\end{table}

\noindent Next we used a simple insight---we noted that npm registry API for individual packages~\cite{registry} often provides the GitHub link (where the code is hosted). This corresponding GitHub page contained the issues raised for a particular package by the developer community (both security and non-security related) as well as discussion on how developers tried to respond to and/or resolve these issues----we collected all of this data. 

\vspace{1mm}

\noindent \textbf{Collecting GitHub links from npm registry}: \newtext{We collected the GitHub links }
for 70,143 randomly sampled packages using the npm registry API. However, not all npm packages contained GitHub links and the same GitHub links occurred in multiple packages---we collected 37,278 distinct GitHub links for 45,466 packages (Table~\ref{npm_packages_stats_1}). 

\vspace{1mm}

\noindent \textbf{Collecting community reported issues data from GitHub}: We used the GitHub API~\cite{GitHub-doc} to collect \textit{all} the issues data for 37,278 distinct GitHub repositories. In total we collected 10,907,467 issues where an overwhelming majority (10,062,759 or 92.3\%) were \textit{closed} issues. For each issue, we collected the issue title, the issue body (i.e., description), and metadata (e.g., the \textit{tags} or usernames which posted issues/comments). We present the summary statistics of our final issues data in Table~\ref{stats_data_issues_collected_1}. Out of these more than 10 million issues, next we identify \textit{security} related issues. 

\subsection{Identifying security-related issues} 

\begin{table}[]
\centering
\scriptsize
\begin{tabular}{|l|l|l|l|}
\hline
\rowcolor[HTML]{9B9B9B} 
\textbf{Tag} &
  \textbf{\begin{tabular}[c]{@{}l@{}}\# repository\\ using the tag\end{tabular}} &
  \textbf{Tag} &
  \textbf{\begin{tabular}[c]{@{}l@{}}\# repository\\ using the tag\end{tabular}} \\ \hline
dependencies  & 7,526 & docs            & 400 \\ \hline
enhancement   & 4,767 & discussion      & 386 \\ \hline
bug           & 3,796 & feature request & 343 \\ \hline
help wanted   & 2,439 & wontfix         & 234 \\ \hline
question      & 1,769 & github\_actions & 197 \\ \hline
good first issue &
  1,079 &
  \cellcolor[HTML]{CBCEFB}\textbf{security} &
  \textbf{196} \\ \hline
documentation & 933   & performance     & 182 \\ \hline
javascript    & 568   & stale           & 169 \\ \hline
feature       & 556   & bug             & 141 \\ \hline
greenkeeper   & 458   & blocked         & 136 \\ \hline
\end{tabular}
\caption{Top 20 most frequently used tags across repositories, along with the number of repositories in which they have been used, highlighting the relative scarcity of security-related tags.}
\label{table:Frequency_of_tags_across_repos}

\end{table}

\noindent GitHub provides the option of tagging each issue\snewtext{~\cite{github_labels,lunny_sane_labels}} where a tag is a small phrase signifying the type of issue. One issue can contain multiple tags. However, each repository can create tags on its own (with arbitrary words). \snewtext{Prior work ~\cite{10.1145/3477314.3507123,JIANG2021106394,7081875,9252051} highlights the effective use of issue tags in OSS for tasks like issue resolution, label prediction, and security related issue identification.} To that end, we started with the idea that perhaps tags which contain security-related words/phrases will be related to security. 

\vspace{1mm}

\noindent\textbf{Tags are used with moderate frequency}: We found that, out of 10,907,467 issues, around 50\% (5,454,149 issues) do not contain any tags. In fact, not all repositories used tags. Out of 37,278 unique repositories, only 13,031 utilized one or more tags. The distribution of tags per repository is shown in Figure \ref{freq_tag_repo} of Appendix \ref{app_sec2}. We collected a total of 23,356 unique tags used across 45,466 npm packages. We show the most popular twenty tags in Table~\ref{table:Frequency_of_tags_across_repos}. \snewtext{We  manually reviewed tags across repositories that are potentially related to security such as ``vulnerability'', ``exploit'', ``cve''; however these tags appeared in fewer than five repositories and did not rank among the top recurring tags.} Interestingly, the word ``security'' appeared as a tag in only 196 repositories.  

\vspace{1mm}
\noindent \textbf{Detecting security-related tags with Word2vec}: Since tags are developer-defined and can be arbitrary, we focused on identifying tags which contain words semantically similar to security. \snewtext{Building on the work of B\"uhlmann et al.~\cite{10.1145/3477314.3507123}, which identified security-related issues by filtering tags containing the term ``security", we created embeddings of each tag in our dataset using Word2Vec and calculated the cosine similarity of those embeddings with the embedding of the term "security". This Word2Vec based tag-detection approach is in line with prior work~\cite{9252051}.} 
We selected tags with similarity of above 0.8 (we tried a number of thresholds and 0.8 gave the most relevant output), resulting in 25 security-related tags (given in Table \ref{tags_sec_w2vec} of Appendix \ref{app_sec2}).
We manually inspected them and found that all contained the word \textit{secure}---overall these tags identified 13,835 potentially security-related issues. \snewtext{We manually checked 100 random such issues and confirmed their relevance to security (e.g., injection risks). Thus, the security-related tags we found (which often contained the substring \textit{security}, e.g., ``security vulnerability''), when present, indicated security-related issues.} However, these security-related issues constitute only 0.13\% of more than 10 million issues reported for these repositories. 

\subsection{Categorizing security-related issues by type of accounts who reported these issues}

\noindent \textbf{Identifying bot-reported security issues}: Bots have become increasingly common on GitHub, automating repetitive and error-prone tasks to facilitate collaborative development~\cite{bot_power}. Bots are, in fact, among the most significant contributors to specific software projects~\cite{bot_recog}. Among the 13,835 security-related issues, 7,731 were created by accounts with \textit{bot} in their usernames, indicating that these issues were likely created by GitHub bots. We further collected 10,500 comments pertaining to these issues and found that 8,442 comments (80.4 \%) are also posted by \newtext{bot} accounts (\newtext{i.e.} with ``bot'' in their username). In total, we identified 93 unique bot accounts; however, the majority (87.6\%) of these issues were created by Dependabot ~\cite{Dependabot2021}. Given the prevalence of bots in creating security-tagged issues we investigate the effectiveness of these bots in addressing security concerns for npm packages in RQ2 (Section~\ref{sec:rq2}). 

\vspace{1mm}

\noindent \textbf{Identifying user-reported security issues}: The remaining 6,104 of the 13,835 issues are created by accounts linked to regular individuals, hereafter referred to as users. Using GitHub API, we further collected a total of 19,324 comments made by accounts while discussing these issues. Out of them only 2,826 (14.6\%) were made by bot accounts. We found 70 more unique bots involved in user-reported issues. Thus our dataset contained a total of 163 bots. However, these bots mostly did not address security-related issues. They primarily posted comments signifying inactivity and staleness of the issue rather than contributing to any meaningful discussion about resolution of the security issues. Next, we focus on these user-reported issues and analyze the \textit{process} of resolution.

\begin{table*}[]
\scriptsize
\centering
\begin{tabular}{l|l|l}
\hline
\rowcolor[HTML]{FFFFFF} 
\multicolumn{1}{|c|}{\cellcolor[HTML]{FFFFFF}\textbf{Creation}}                            & \multicolumn{1}{c|}{\cellcolor[HTML]{FFFFFF}\textbf{Discussion}}                                     & \multicolumn{1}{c|}{\cellcolor[HTML]{FFFFFF}\textbf{Resolution}}                             \\ \hline
\rowcolor[HTML]{CBCEFB} 
\multicolumn{1}{|l|}{\cellcolor[HTML]{CBCEFB}A. Issue with solution(PR)}                   & A. Acknowledged                                                                                      & \multicolumn{1}{l|}{\cellcolor[HTML]{CBCEFB}A. With valid reason}                            \\ \hline
\rowcolor[HTML]{C0C0C0} 
\multicolumn{1}{|l|}{\cellcolor[HTML]{C0C0C0}A.1 Description present}                      & A.1 Spoke against with issue                                                                         & \multicolumn{1}{l|}{\cellcolor[HTML]{C0C0C0}A.1 Falsely Created}                             \\ \hline
\rowcolor[HTML]{FFFFFF} 
\multicolumn{1}{|l|}{\cellcolor[HTML]{FFFFFF}A.1.1 Description of the PR}                  & A.1.1 Duplicate of another issue                                                                     & \multicolumn{1}{l|}{\cellcolor[HTML]{FFFFFF}A.1.1 Completed - False Positive}                \\ \hline
\rowcolor[HTML]{FFFFFF} 
\multicolumn{1}{|l|}{\cellcolor[HTML]{FFFFFF}A.1.2 Description with testing instructions}  & A.1.3 Rejected the issue (eg: false positive)                                        & \multicolumn{1}{l|}{\cellcolor[HTML]{FFFFFF}A.1.2 Re-reporting of previously reported issue} \\ \hline
\rowcolor[HTML]{C0C0C0} 
\multicolumn{1}{|l|}{\cellcolor[HTML]{C0C0C0}A.2  No description}                          & \cellcolor[HTML]{FFFFFF}A.1.4. Policy adherence & \multicolumn{1}{l|}{\cellcolor[HTML]{C0C0C0}A.2 Successfully resolved}                       \\ \hline
\multicolumn{1}{|l|}{\cellcolor[HTML]{CBCEFB}B. Issue without solution}                    & \cellcolor[HTML]{C0C0C0}A.2 Spoke for the issue                                                      & \multicolumn{1}{l|}{\cellcolor[HTML]{FFFFFF}A.2.1 Issue resolved in discussion}              \\ \hline
\rowcolor[HTML]{FFFFFF} 
\multicolumn{1}{|l|}{\cellcolor[HTML]{C0C0C0}B.1 Reproducibility}                          & A.2.1 Discussing about problem                                                                       & \multicolumn{1}{l|}{\cellcolor[HTML]{FFFFFF}A.2.2 Completed by merging PR}                   \\ \hline
\rowcolor[HTML]{FFFFFF} 
\multicolumn{1}{|l|}{\cellcolor[HTML]{FFFFFF}B.1.1 Description with code snippet}          & \cellcolor[HTML]{e4ebd4}A.2.1.1 Faced same problem                                                                           & \multicolumn{1}{l|}{\cellcolor[HTML]{FFFFFF}A.2.3 Refer to other PR/commit}                  \\ \hline
\rowcolor[HTML]{FFFFFF} 
\multicolumn{1}{|l|}{\cellcolor[HTML]{FFFFFF}B.1.2 Description with steps of reproduction} & \cellcolor[HTML]{e4ebd4}A.2.1.2 Accepted the issue                                                                           & \multicolumn{1}{l|}{\cellcolor[HTML]{C0C0C0}A.3 To be completed}                             \\ \hline
\rowcolor[HTML]{FFFFFF} 
\multicolumn{1}{|l|}{\cellcolor[HTML]{FFFFFF}B.1.3 Description with error logs}            & \cellcolor[HTML]{e4ebd4}A.2.1.3 Able to reproduce                                                                            & \multicolumn{1}{l|}{\cellcolor[HTML]{FFFFFF}A.3.1 Deferred fix (next version)}               \\ \hline
\rowcolor[HTML]{FFFFFF} 
\multicolumn{1}{|l|}{\cellcolor[HTML]{FFFFFF}B.1.4 Description with system info}           & \cellcolor[HTML]{e4ebd4}A.2.1.4. Asking for more clarification                                               & \multicolumn{1}{l|}{\cellcolor[HTML]{FFFFFF}A.3.2 Won't fix / Can’t solve now}               \\ \hline
\multicolumn{1}{|l|}{\cellcolor[HTML]{C0C0C0}B.2 Non-reproducibility}                      & \cellcolor[HTML]{FFFFFF}A.2.2 Discussing about solution                                              & \multicolumn{1}{l|}{\cellcolor[HTML]{CBCEFB}B. Without valid reason}                         \\ \hline
\rowcolor[HTML]{FFFFFF} 
\multicolumn{1}{|l|}{\cellcolor[HTML]{FFFFFF}B.2.1 No Description}                         & \cellcolor[HTML]{e4ebd4}A.2.2.1 Solved the issue in the discussion                                                           & \multicolumn{1}{l|}{\cellcolor[HTML]{C0C0C0}B.1 Closed without reason}                       \\ \hline
\rowcolor[HTML]{FFFFFF} 
\multicolumn{1}{|l|}{\cellcolor[HTML]{FFFFFF}B.2.2 Only Description}                       & \cellcolor[HTML]{e4ebd4}A.2.2.2 Suggested changes/solution                                                                   & \multicolumn{1}{l|}{\cellcolor[HTML]{C0C0C0}B.2 Completed due to staleness}                  \\ \hline
\rowcolor[HTML]{FFFFFF} 
\multicolumn{1}{|l|}{\cellcolor[HTML]{FFFFFF}B.2.3 Feature Requests}                       & \cellcolor[HTML]{e4ebd4}A.2.2.3 Asked for corrections                                                                        &                                                                                              \\ \hhline{|-|-|} 
\rowcolor[HTML]{FFFFFF} 
               & \cellcolor[HTML]{e4ebd4}A.2.2.4 Agree with the solution                                                                      &                                                                                              \\ \hhline{|~|-|}
\rowcolor[HTML]{FFFFFF} 
                                                                                           & \cellcolor[HTML]{e4ebd4}A.2.2.5 Disagree with the given solution                                                             &                                                                                              \\ \hhline{|~|-|}
\rowcolor[HTML]{FFFFFF} 
                                                                                           & \cellcolor[HTML]{CBCEFB}B Ignored                                                                    &                                                                                              \\ \hhline{|~|-|}
\rowcolor[HTML]{FFFFFF} 
                                                                                           & \cellcolor[HTML]{C0C0C0}B.1 Not interested                                                           &                                                                                              \\ \hhline{|~|-|}
\rowcolor[HTML]{FFFFFF} 
                                                                                           & B.1.1 Doesn't encourage solving the problem                                                          &                                                                                              \\ \hhline{|~|-|}
\rowcolor[HTML]{FFFFFF} 
                                                                                           & B.1.2 No comments                                                                                    &                                                                                              \\ \hhline{|~|-|}
\rowcolor[HTML]{FFFFFF} 
                                                                                           & \cellcolor[HTML]{C0C0C0}B.2 Inconclusive                                                             &                                                                                              \\ \cline{2-2}
\end{tabular}
\caption{Our four-level hierarchical themes explaining the action/interaction between users/developers for security related issues.}
\label{four_level_hierarchy}
\end{table*}

\section{Uncovering Process of Resolving User-Reported Security Issues in npm Using Qualitative Analysis}\label{sec:affinity_coding}

\noindent We next asked: how user-reported security issues are created, discussed and resolved for npm packages. 

\vspace{1mm}

\noindent \textbf{Extraction and division of quotes:} We identified three prominent phases in the life-cycle of each security issue: Creation phase (captured by the user-reported description), discussion phase (interaction between developers of GitHub repositories for npm packages and users),  resolution phase (captured by the last comment before the closing event). We first take 6,104 user-reported security issues as well as 16,498 comments and 6,104 closing events from our dataset. We programmatically divided this data into three phases and from each phase randomly selected 500 issue body/comments. Two researchers together extracted a total of 1,747 explanatory quotes from these issue body/comments across three phases. 
Next, we use open coding and affinity diagramming~\cite{saldana2015} to develop a hierarchy of themes explaining user/developer action/interaction for security-related issues in each of the three phases.

\vspace{1mm}

\noindent \textbf{Open coding:} Initially, we open-coded the quotes. For each of the three phases, we randomly selected 100 quotes and then two researchers cooperatively developed three codebooks. The codebook from the creation phase captures the types of information provided by a user, for the discussion phase, it captures how npm package developers responded to security issues, and for the resolution phase, it captures if there is any successful resolution. Additionally, we set aside 25 quotes initially to assess the saturation of themes. \newtext{Specifically, we started coding 100 ($\sim$5\% of) quotes (in line with Raj et al.~\cite{298100})}. Subsequently, the two researchers used the codebooks to independently code all the quotes in each phase. Inter-rater agreement \newtext{(Cohen’s Kappa)} at the end of the open coding round was 0.85, indicating almost perfect agreement. At the end of open coding, the two researchers met to discuss and resolve the disagreements. Then one researcher verified that the resultant codes were sufficient to code the 25 quotes, indicating thematic saturation\newtext{~\cite{saunders2018saturation}.} Finally, we ended up with a total of 13 codes across three phases. 

\vspace{1mm}

\noindent \textbf{Affinity diagramming to identify the patterns in discussing security-related issues:} After the open coding round, the two researchers used affinity diagramming to jointly examine the discovered codes. They did this by looking at the collection of quotes for each code in addition to the code itself~\cite{10.1145/2702123.2702561}. We set aside 10 random codes to check for saturation at the end. Then the coders collaboratively created higher-level themes from the rest of the codes. They kept doing this with the new higher-level themes for two more rounds, or until the coders thought that no more new higher-level themes could come up. In the end, we ended up with a four-level hierarchy of themes, capturing the process of creation-discussion-resolution of user-reported security issues in npm packages. Level 1 themes encompassed abstract, overarching themes generated in the final round of affinity diagramming, and Level 4 comprised the individual codes established during the open coding phase. Finally, we checked that including the 10 random codes did not add any new Level-1 and Level-2 themes, indicating thematic saturation of affinity diagramming. Our hierarchy of themes (first four levels) explaining the patterns in the resolution of user-reported security-related issues is shown in Table \ref{four_level_hierarchy}. However, we note that so far we only leveraged 6,104 user-reported issues, which limits our dataset and analysis. 

\section{Scaling User-Reported Security Issues Dataset and Theme Annotation}\label{sec:scaling}

\begin{figure}
    \centering
\includegraphics[width=1\linewidth]{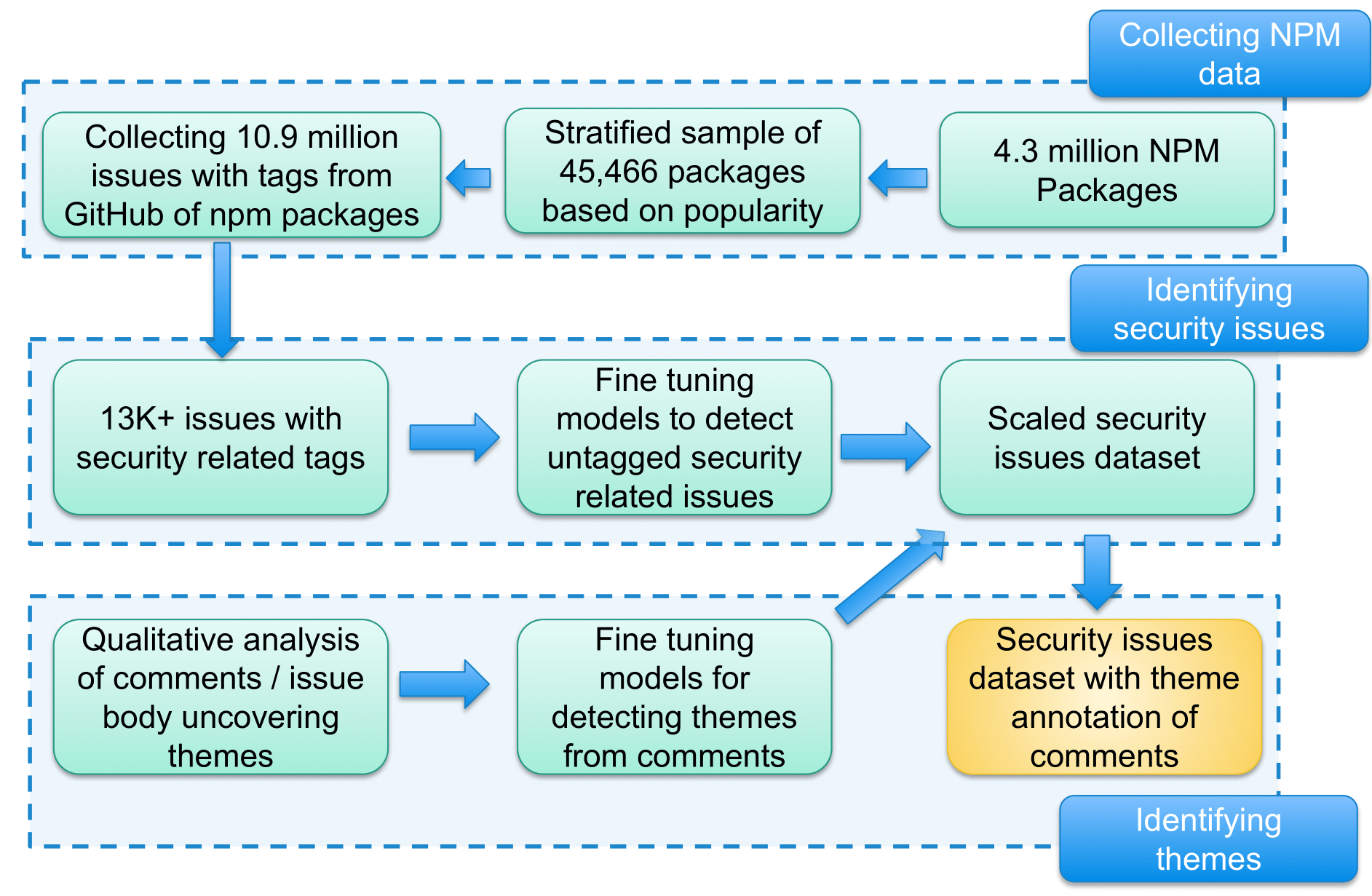}
    \caption{Pipeline for filtering security issues and identifying themes occurring in them.}
    \label{flowchart}
\end{figure}

\noindent Our dataset of security-related issues contains only 6,104 issues (0.13\% of a total of 10,907,467 issues). To that end, we note a potential reason: although bot-reported issues are tagged automatically, for user-reported issues, the tags are assigned by npm package developers. Thus, it might be that there are security-related issues which are not tagged with security-related tags. Thus, to identify such user-reported issues and scale up our dataset, we designed a machine learning-based pipeline presented in \newtext{Figure \ref{flowchart}}. 

\subsection{Leveraging text classification to extend the security-related issue dataset}

\noindent \textbf{Creating ground truth labels}: We view the problem \newtext{of} detecting security-related issues within the set of issues without any security-related tags as a classification task. Thus, we manually create a gold-standard dataset of 2,000 issues with two classes (security-related and non-security-related) as follows. We randomly sample the issue description of 1000 issues from the set of 6,104 user-reported issues with security-related tags. Furthermore, we randomly sampled 1000 issues which did not have security-related tags. Then two coders code these 1000 issues independently using two labels--- security related or non-security related. The Cohen's kappa for the two coders was 0.74, showing substantial agreement. Then they meet, resolved disagreements, and assign final labels. \newtext{To handle cases where issues had both security-related and non-security-related tags, we applied a simple rule: if an issue had at least one security-related tag, we labeled it as a security issue.} In total, combining these two steps, we ended up with 1042 security-related and 1058 non-security-related issues. 

\begin{table}[]
\scriptsize
\centering
\begin{tabular}{|c|c|c|}
\hline
\rowcolor[HTML]{C0C0C0}
\textbf{Model}   & \textbf{Accuracy} & \textbf{Macro avg F1} \\ \hline
CodeBERT         & 0.9               & 0.9                   \\ \hline
BERT             & 0.93              & 0.93                  \\ \hline
\textbf{RoBERTa} & \textbf{0.94}     & \textbf{0.94}         \\ \hline
FLAN T5          & 0.89              & 0.89                  \\ \hline
DeBERTa          & 0.94              & 0.94                  \\ \hline
\end{tabular}
\caption{Performance of our fine-tuned models on the test dataset for classifying issues into security-related and non-security-related.}
\label{Performance_finetuned_bertmodels}
\end{table}

\noindent
We model the problem of discovering potential security-related issues (which are not tagged with security-related tags) as a two-class classification problem (where the issue title and issue description will be used to classify the text). To that end, we experimented with various models as classifiers, including BERT-based models and large language models (LLMs), the latter in both zero-shot and few-shot in-context learning (ICL) settings. Among these, we selected fine-tuned RoBERTa for further analysis, as it demonstrated the best performance (F1-score of 0.94) (detailed performance metrics for the BERT-based models are provided in Table \ref{Performance_finetuned_bertmodels}. LLMs performed worse than RoBERTa, achieving an F1 score above 80\%---the results are given in \newtext{Appendix~\ref{sec-issu}}, Table \ref{Performance_finetuned_llms}).

\noindent
\textbf{Model Architecture:} Given the nature of our task and the availability of labelled data, we utilized a range of pretrained models to address the classification challenge. These included BERT~\cite{devlin2018bert}, RoBERTa~\cite{liu2019roberta}, CodeBERT~\cite{feng2020codebert}, FLAN-T5~\cite{chung2022scalinginstructionfinetunedlanguagemodels} and DeBERTa~\cite{he2020deberta}, all of which were fine-tuned on our dataset for the downstream task of issues classification. In addition, we evaluated LLMs, namely Mistral~\cite{jiang2023mistral}, Qwen~\cite{qwen}, Meta-LLaMA~\cite{touvron2023llama}, and Gemma~\cite{team2024gemma} using zero-shot and few-shot under ICL settings. We leveraged systematic hyperparameter tuning using the Optuna framework ~\cite{akiba2019optunanextgenerationhyperparameteroptimization} to ensure best model configurations. 

Given the sequence classification nature of our task, we utilized a range of model architectures, as outlined in Table \ref{details_bertmodels_arch} of \newtext{Appendix~\ref{sec-issu}}. We achieved the best accuracy for the downstream task using the fine-tuned RoBERTa model---class-wise performance of the RoBERTa model is shown in Table \ref{Roberta_classwise_performance}. Although DeBERTa showed similar performance on the test dataset, the manual analysis revealed it produced numerous false positives and negatives, hinting at its unsuitability (misclassification analysis of DeBERTa is in Table~\ref{misclassification_Debarta} of \newtext{Appendix~\ref{sec-issu}}).

\noindent \textbf{Model Performance:} We randomly split the dataset into 80\%
training and 20\% validation subsets. Post-training, the RoBERTa model achieved an accuracy of 94\% and an F1-score of 94\% \snewtext{and a ROC-AUC score of 0.94} (details are provided in Table \ref{Performance_finetuned_bertmodels}). To further investigate the performance and misclassification of our model, we randomly selected an additional 100 issues (excluding the \newtext{1,600} issues used for training) and one annotator manually annotated the selected issues to create ground truth. Next, we used our text classification on these 100 reviews and constructed the confusion matrix (Table \ref{Miscallsification_roberta}). \newtext{We note that, in this confusion matrix, only 4 issues were misclassified between security and non-security—three security issues as non-security and one non-security issue as security.}
Thus, we identify security-related issues correctly in 96\% of cases. 
\snewtext{Our classifiers were trained on issue body and title marked with security-related tags–-out of 1,000 classifier-flagged randomly sampled security-related issues, 57\% didn’t contain the string “security”. To further validate the model’s performance, we extracted attention scores from the [CLS] token across all heads in the final RoBERTa layer, averaged them, and visualized the most influential tokens using a word cloud shown in Figure- \ref{fig:wordcloud} (Appendix ~\ref{sec-issu})~\cite{aditya2021cls}.}

\begin{table}[]
\scriptsize
\centering
\begin{tabular}{llll}
\cline{3-4}
                                                                                                        & \multicolumn{1}{l|}{}             & \multicolumn{2}{c|}{\textbf{Predicted}}                                                         \\ \cline{3-4} 
                                                                                                        & \multicolumn{1}{l|}{}             & \multicolumn{1}{l|}{Security}                  & \multicolumn{1}{l|}{Non Security}              \\ \hline
\multicolumn{1}{|c|}{}                                                                                  & \multicolumn{1}{l|}{Security}     & \multicolumn{1}{l|}{6}                         & \multicolumn{1}{l|}{\cellcolor[HTML]{C0C0C0}3} \\ \cline{2-4} 
\multicolumn{1}{|c|}{\multirow{-2}{*}{\textbf{\begin{tabular}[c]{@{}c@{}}Ground\\ Truth\end{tabular}}}} & \multicolumn{1}{l|}{Non Security} & \multicolumn{1}{l|}{\cellcolor[HTML]{C0C0C0}1} & \multicolumn{1}{l|}{90}                        \\ \hline
                                                                                                        &                                   &                                                &                                                \\
                                                                                                        &                                   &                                                &                                               
\end{tabular}
\caption{ Confusion matrix for our RoBERTa text-classifier annotation and ground truth for previously unseen 100 issues. 
\newtext{Only four cases were misclassified between the security and non-security categories} (marked in gray).}
\label{Miscallsification_roberta}
\end{table}

\vspace{1mm}

\noindent \textbf{Applying the model to scale up user-reported untagged security issues:} Finally, we used our validated issue-classifier model on 9,131,800 user-reported issues which were not tagged with any security-related tags.
Our model identified 1,617,738 issues as potentially security-related (14.8\% of all issues)--- representing a significant rise from the originally tagged security issues, which uncovered only 13,835 security-related issues (0.13\% of all issues). Thus, our work uncovered 114 times more security-related issues reported by users than actually tagged by npm package developers. A few examples of these issues are provided in Table \ref{examples_5_5_issues} in Appendix \ref{sec-issu}. This surprising result emphasized the substantial volume of security issues that remain untagged across various GitHub repositories of npm packages. 

\subsection{Identifying themes mentioned in extended dataset}

\noindent Next, we aimed to attribute themes to the quotes in our extended dataset of 1,617,738 potentially security-related issues. To that end, we collected a total of 4,461,934 comments and issue bodies and assigned themes to them using a similar idea as before---we modelled this theme identification problem \newtext{as a} multi-label classification task. In particular, we represent the Level-2 \newtext{(L-2)} themes as labels that are allocated to each quote. We opted to conduct our study using L-2 themes, as L-1 themes were deemed excessively wide for detailed examination. Also, L-3 and L-4 themes might be more granular than L-2 themes, but we have a relatively low amount of labelled data per L-3/L-4 theme. Thus, automated theme identification with high accuracy would have been very difficult for them. 

\vspace{1mm}

\noindent \textbf{Model Architecture:} We again used the following pre-trained models, BERT~\cite{devlin2018bert}, RoBERTa~\cite{liu2019roberta}, CodeBERT~\cite{feng2020codebert}, FLAN-T5~\cite{chung2022scalinginstructionfinetunedlanguagemodels} and DeBERTa~\cite{he2020deberta}. Given the different
architectures of these models, we fine-tuned BERT with multi-head classification layers on top of the pre-trained model. FLAN-T5 were fine-tuned on the seq2seq task. We used the Optuna framework ~\cite{akiba2019optunanextgenerationhyperparameteroptimization} for systematic hyperparameter tuning.

\vspace{1mm}

\noindent \textbf{Model Performance:} We divided the labelled dataset into 80\% for training and 20\% for validation to assess the performance of the fine-tuned models. We found that the RoBERTa model outperformed other models, identifying more than 80\% of themes correctly in the reviews. The results of each model's accuracy on the validation set are displayed in Table \ref{model_stats_theme_identification}. In order to further verify the accuracy of the model, we conducted a manual analysis. Specifically, we selected 100 random quotations (excluding the original 1,638 annotated) and manually annotated them with Level-2 themes. Then, we compared the correctness of our predicted themes against this ground truth. The result of the manual analysis is detailed in Table \ref{table_themes_roberta_appendix} (Appendix \ref{app_theme}). In both the validation set and the additional ground truth dataset, our classifier was accurate for the majority of the evaluations (almost 80\%). Next, we will leverage this extended dataset of theme-annotated issues to explore our research questions.

\begin{table}[]
\scriptsize
\centering
\begin{tabular}{|l|l|l|}
\hline
\rowcolor[HTML]{C0C0C0} 
{\color[HTML]{000000} \textbf{Model}}   & {\color[HTML]{000000} \textbf{Accuracy}} & {\color[HTML]{000000} \textbf{\begin{tabular}[c]{@{}l@{}}Macro avg\\ F1\end{tabular}}} \\ \hline
{\color[HTML]{000000} CodeBERT}         & {\color[HTML]{000000} 0.78}              & {\color[HTML]{000000} 0.67}                                                            \\ \hline
{\color[HTML]{000000} BERT}             & {\color[HTML]{000000} 0.77}              & {\color[HTML]{000000} 0.63}                                                            \\ \hline
{\color[HTML]{000000} \textbf{RoBERTa}} & {\color[HTML]{000000} \textbf{0.82}}     & {\color[HTML]{000000} \textbf{0.79}}                                                   \\ \hline
{\color[HTML]{000000} FLAN T5}          & {\color[HTML]{000000} 0.65}               & {\color[HTML]{000000} 0.44}                                                            \\ \hline
{\color[HTML]{000000} DeBERTa}          & {\color[HTML]{000000} 0.80}               & {\color[HTML]{000000} 0.75}                                                            \\ \hline
\end{tabular}
\caption{Performance of fine-tuned RoBERTa model (for identification of themes explaining the action/interaction between users/developer) on the validation dataset.}
\label{model_stats_theme_identification}
\end{table}

After validating the model on both the validation set and an additional ground truth dataset—where it achieved nearly 80\% accuracy across most evaluations—we applied it to identify themes within the entire dataset of 1,617,738 issues. A detailed summary of the themes identified is provided in Tables \ref{1_5M_inferance_creation}, \ref{1_5M_inferance_discussion}, and \ref{1_5M_inferance_resolution} in Appendix \ref{app_theme}. Additionally, a summary of the number of items of each class is provided in Table \ref{model_pred_1_5M_L2}.

\begin{table}[]
\scriptsize
\centering
\begin{tabular}{|l|l|l|l|}
\hline
\rowcolor[HTML]{C0C0C0}   & \textbf{Precison} & \textbf{Recall} & \textbf{F1 score} \\ \hline
\cellcolor[HTML]{C0C0C0}\textbf{\begin{tabular}[c]{@{}l@{}}Security Related\\ Issue\end{tabular}}     & 0.95              & 0.93            & 0.94              \\ \hline
\cellcolor[HTML]{C0C0C0}\textbf{\begin{tabular}[c]{@{}l@{}}Non Security\\ Related Issue\end{tabular}} & 0.93              & 0.95            & 0.94              \\ \hline
\end{tabular}
\caption{Class-wise performance of our fine-tuned RoBERTa model on the validation dataset.}
\label{Roberta_classwise_performance}
\end{table}

\begin{table}[]
\scriptsize
\centering
\begin{tabular}{|l|l|l|}
\hline
\rowcolor[HTML]{C0C0C0} 
\textbf{Type}                         & \textbf{L2 levels}         & \textbf{Count} \\ \hline
                                      & Description present        & 786,818        \\ \cline{2-3} 
                                      & No description            & 39,212         \\ \cline{2-3} 
                                      & Reproducibility            & 418,668        \\ \cline{2-3} 
\multirow{-4}{*}{\textbf{Creation}}   & Non-reproducibility        & 372,700        \\ \hline
                                      & Spoke against with issue   & 194,971        \\ \cline{2-3} 
                                      & Spoke for the issue        & 1,786,539      \\ \cline{2-3} 
                                      & Not interested             & 133,035        \\ \cline{2-3} 
\multirow{-4}{*}{\textbf{Discussion}} & Inconclusive               & 353,046        \\ \hline
                                      & Falsely Created            & 73,670         \\ \cline{2-3} 
                                      & Successfully resolved      & 727,861        \\ \cline{2-3} 
                                      & To be completed            & 480,391        \\ \cline{2-3} 
                                      & Closed without reason     & 79,188         \\ \cline{2-3} 
\multirow{-5}{*}{\textbf{Resolution}} & Completed due to staleness & 37,230         \\ \hline
\end{tabular}
\caption{Breakdown of activities across creation, discussion, and resolution phases, detailing subcategories (L2 levels) and their respective counts as identified by our fine-tuned RoBERTa model.}
\label{model_pred_1_5M_L2}
\end{table}

\section{How Prevalent is Reporting Security-Related Issues in GitHub Repositories of npm Packages? (RQ1)}\label{sec:rq1}

\noindent Using our machine learning model, we identified more than 1.6 million (14.8\%) previously untagged security-related issues within the dataset. We further explore whether these issues are evenly distributed across repositories, the extent of community discussions surrounding them, the time taken to resolve them, and the presence of CVE IDs in these reports.

\begin{table}[]
\scriptsize
\centering
\begin{tabular}{|l|l|l|}
\hline
\rowcolor[HTML]{C0C0C0} 
\textbf{\begin{tabular}[c]{@{}l@{}}\# dependent \\ packages\end{tabular}} & \textbf{\begin{tabular}[c]{@{}l@{}}\% GitHub repositories that \\ had at least one un-tagged\\ security issue\end{tabular}} & \textbf{\begin{tabular}[c]{@{}l@{}}\% detected issues that\\ were un-tagged\end{tabular}} \\ \hline
0 & 16.03 & 15.88 \\ \hline
0-10 & 24.75 & 16.66 \\ \hline
10-100 & 39.34 & 14.54 \\ \hline
100-500 & 44.3 & 14.72 \\ \hline
500-1000 & 55.77 & 10.94 \\ \hline
\textgreater{}1000 & 66.09 & 10.98 \\ \hline
\end{tabular}
\caption{Distribution of security-related issues across \\ different buckets.}
\label{distribution_1_5M}
\end{table}

\vspace{1mm}

\noindent \textbf{How are these security-related issues distributed over the buckets?} We first check the fraction of issues reported in our buckets of npm repositories based on their dependent repositories. The result is presented in Table~\ref{distribution_1_5M}. The percentage of repositories with security-related issues increased as we moved to buckets with npm packages with more dependents. However, the percentage of security-related issues remained fairly consistent across all buckets, indicating that the overall occurrence of such issues is relatively uniform over buckets. 

\vspace{1mm}

\noindent \textbf{How much do the community discuss about security-related issues? }We examined user engagement by analyzing comments on security-related issues. Our dataset of 1,617,738 security issues, spread across 15,048 repositories, contains a total of 4,461,934 comments. More than 23\% of the issues do not even receive any comments highlighting the lack of interaction in these security-related issues. 

\vspace{1mm}

\noindent \textbf{Do security-related issues persist over time and take longer to resolve? }We observed that security-related issues appeared consistently over the years, highlighting their persistent nature across different periods (Table~\ref{table:Issues_created_back_ago} of Appendix \ref{app_rq1}).To evaluate whether security-related issues require more time to resolve, we computed the resolution time (difference between its opening and closing timestamps) for a random sample of 10,000 non-security-related issues and compared it with the resolution time of security-related issues. Security-related issues take an average of 56.47 days to close, compared to an average of 48.88 days for non-security issues.

\vspace{1mm}

\noindent \textbf{Do these issues contain specific mention of CVE IDs? }Among the security-related issues, only 4,783 issues (out of 1.6 million) mention a total of 7,184 CVE IDs, with 2,233 of these being unique. Thus the low inclusion of \newtext{CVE ID} highlights that publicly recognized vulnerabilities might not help the users to identify the security concerns within the npm ecosystem. Thus within GitHub repositories of npm packages, numerous security issues have not been formally identified or documented under CVE IDs. \newtext{(Note: This analysis considers only CVE mentions.)} The five most frequently mentioned CVE IDs are in Table ~\ref{cve_5_description} of Appendix \ref{app_rq1}.

\section{Effectiveness of Bots in Detecting and Addressing Security Related Issues (RQ2)}\label{sec:rq2}

Next, we explore how bots function within GitHub repositories of npm packages since out of 13,835 issues with security-related tags, 7,731 (55.9\%) were bot-reported. 

\subsection{Understanding role of bots in bot-reported security-related issues in npm}

In Section~\ref{sec:data_creation} we noted that Dependabot reported 87.6\% bot-reported issues in GitHub repositories of npm packages. It is an automated tool that manages project dependencies by identifying and updating vulnerable packages (among other functions). Dependabot works by analyzing dependency files, including package.json and  package-lock.json, to check for outdated or insecure packages. When a vulnerability is detected, Dependabot checks the lock file to identify the affected version and its impact on other dependencies. It provides two main services-- 1) version updates, which automatically update dependencies to the latest versions according to the configuration in \textit{dependabot.yml}, and 2) security updates, which scan repositories for known vulnerabilities and alert repository owners. 

Despite its utility, Dependabot can overwhelm developers with numerous pull requests, especially in repositories with many dependencies, making management difficult~\cite{dependabot_hackernews_2, dependabot_hackernews}. It often creates excessive pull requests for bloated dependencies, and its configuration can cause issues for developers. Additionally, Dependabot's reliance on predefined vulnerability databases limits its ability to detect undisclosed or emerging threats(e.g.: supply-chain attacks~\cite{10.1007/978-3-030-52683-2_2}), leaving security gaps.

\vspace{1mm}

\begin{table*}[]
\centering
\scriptsize
\begin{tabular}{|p{3.5cm}|p{4cm}|p{3.5cm}|}
\hline
\cellcolor[HTML]{C0C0C0}\textbf{A. Dependency management} & \cellcolor[HTML]{FFFFFF}github-actions{}               & bors{}                                           \\ \hline
\cellcolor[HTML]{FFFFFF}dependabot                        & \cellcolor[HTML]{C0C0C0}\textbf{D. Security}                  & \cellcolor[HTML]{FFFFFF}github-merge-queue{}     \\ \hline
\rowcolor[HTML]{FFFFFF} 
renovate{}                                         & socket-security{}                                      & \cellcolor[HTML]{CBCEFB}\textbf{E.3 Project Management} \\ \hline
\cellcolor[HTML]{C0C0C0}\textbf{B. PR management}         & \cellcolor[HTML]{FFFFFF}robocop{}                      & \cellcolor[HTML]{FFFFFF}issue-sh{}               \\ \hline
\rowcolor[HTML]{FFFFFF} 
stale{}                                            & mend-for-github-com{}                                  & \cellcolor[HTML]{FFFFFF}angular-robot{}          \\ \hline
\rowcolor[HTML]{FFFFFF} 
mergify{}                                          & github-advanced-security{}                             & \cellcolor[HTML]{FFFFFF}project-bot{}            \\ \hline
\cellcolor[HTML]{FFFFFF}pullapprove{}              & \cellcolor[HTML]{C0C0C0}\textbf{E. Utility}                   & \cellcolor[HTML]{FFFFFF}dosubot{}                \\ \hline
\cellcolor[HTML]{FFFFFF}delete-merged-branch{}     & \cellcolor[HTML]{CBCEFB}\textbf{E.1 Policy \& CLA}            & \cellcolor[HTML]{FFFFFF}sync-by-unito{}          \\ \hline
\rowcolor[HTML]{FFFFFF} 
coderabbitai{}                                     & microsoft-github-policy-service{}                      & \cellcolor[HTML]{CBCEFB}\textbf{E.4 Issue Related}      \\ \hline
\rowcolor[HTML]{FFFFFF} 
kodiakhq{}                                         & salesforce-cla{}                                       & \cellcolor[HTML]{FFFFFF}git2gus{}                \\ \hline
\rowcolor[HTML]{FFFFFF} 
trunk-io{}                                         & google-cla{}                                           & \cellcolor[HTML]{FFFFFF}dotnet-issue-labeler{}   \\ \hline
\rowcolor[HTML]{FFFFFF} 
gitpod-io{}                                        & dotnet-policy-service{}                                & \cellcolor[HTML]{FFFFFF}linear{}                 \\ \hline
\cellcolor[HTML]{C0C0C0}\textbf{C. CI/CD}                 & \cellcolor[HTML]{FFFFFF}linux-foundation-easycla{}     & \cellcolor[HTML]{CBCEFB}\textbf{E.5 Code Coverage}      \\ \hline
\cellcolor[HTML]{FFFFFF}cypress{}                  & \cellcolor[HTML]{CBCEFB}\textbf{E.2 PR related}               & \cellcolor[HTML]{FFFFFF}codecov {}               \\ \hline
\rowcolor[HTML]{FFFFFF} 
codesandbox-ci{}                                   & \cellcolor[HTML]{FFFFFF}changeset-bot{}                & codeclimate{}                                    \\ \hline
\cellcolor[HTML]{FFFFFF}azure-pipelines{}          & \cellcolor[HTML]{FFFFFF}jupyterlab-probot{}            & \cellcolor[HTML]{CBCEFB}\textbf{E.6 Miscellaneous
Utility}       \\ \hline
\rowcolor[HTML]{FFFFFF} 
shields-deployment{}                               & \cellcolor[HTML]{FFFFFF}jupyterlab-dev-mode{}          & lock{}                                           \\ \hline
\rowcolor[HTML]{FFFFFF} 
vercel{}                                           & \cellcolor[HTML]{FFFFFF}pull{}                         & gitpoap-bot{}                                    \\ \hline
\rowcolor[HTML]{FFFFFF} 
nx-cloud{}                                         & \cellcolor[HTML]{FFFFFF}conventional-commit-lint-gcf{} & welcome{}                                        \\ \hline
\rowcolor[HTML]{FFFFFF} 
netlify{}                                          & \cellcolor[HTML]{FFFFFF}pull-request-size{}            & gitwave{}                                        \\ \hline
\rowcolor[HTML]{FFFFFF} 
render{}                                           & \cellcolor[HTML]{FFFFFF}release-clerk{}                & sentry-io{}                                      \\ \hline
\cellcolor[HTML]{FFFFFF}gatsby-cloud{}             & \cellcolor[HTML]{FFFFFF}what-the-diff{}                & \cellcolor[HTML]{C0C0C0}\textbf{F. Miscellaneous}                \\ \hline
\rowcolor[HTML]{FFFFFF} 
grafana-delivery-bot{}                             & \cellcolor[HTML]{FFFFFF}a8c-probot-stale{}             & sonarcloud{}                                     \\ \hline
\end{tabular}
\caption{Hierarchical levels of functionalities of bots(along with the bots).}
\label{bots_affinity_diagram}
\end{table*}

\subsection{Understanding role of bots in user-reported security-related issues in npm}
Apart from actively creating more than seven thousand issues tagged as security, bots are also involved in user-reported issues (with account names containing the phrase ``bot''). The activities along with their frequencies are given in Table \ref{table:Frequency_of_botactivities}, Appendix~\ref{app_rq2}. We found that 120 bots were involved in such issues. We focus on them for further understanding their effectiveness in reporting and mitigating security issues in GitHub repositories of npm packages. 

\vspace{1mm}
\noindent \newtext{\textbf{Classification of bots}}
We found that many of these bots are hosted on the GitHub Marketplace ~\cite{GitHubMarketplace2025}, which may provide free or paid access to the users, while others are independently created by developers. However, hosting a bot on the GitHub Marketplace does not guarantee the availability of its source code publicly; one can keep its source code private. Moreover, some bots are restricted within its own organisation. Based on these observations, we classified the 120 bots into two main categories:
(See Table \ref{tab:bot_classification_usercreated_issues} of Appendix \ref{app_rq2})

\vspace{1mm}
\noindent \textbf{Bots deployed on GitHub marketplace}\label{sec_5_2_2}: We found 71 bots which are deployed on GitHub marketplace. These can be further classified based on the availability of their source codes. We identified 57 bots which are hosted for public usage on the GitHub marketplace.  40 of these bots have their source codes publicly available, while 17 of them preferred to keep their source codes private. Their transparency allows developers to review their implementation, modify them for specific needs, and integrate them into their workflows effectively. The remaining 14 bots of the 71 are those whose source codes and accessibility are kept private. These are mostly used internally by the organisations that created them.

\vspace{1mm}

\noindent \textbf{Bots not deployed on GitHub marketplace}: The remaining 49 bots were not deployed in GitHub marketplace. Moreover, they are privately maintained by individuals or organisations. 
These accounts were recognized because their usernames included the term \newtext{``bot''}. 
Unlike GitHub apps, these are not standardized, making it unclear whether they are real bots or simply user accounts named as such. The next subsection discusses their behaviour and roles in more detail.

\subsection{Uncovering functionality of bots}
To better understand the role these bots play in security-related issues, we analyzed various activities that bots perform (using publicly available information). We considered the bots from two groups: those with publicly available source codes and those with private repositories but listed in the marketplace, providing descriptions about their working for analysis. Combining both the categories, we analyze 51 bots. 

\noindent \subsubsection{Qualitative analysis of bot functionalities}
For each of the 51 bots, we collected their GitHub repositories (readme, comments) and marketplace descriptions. We extracted explanatory quotes about bot functionality from these data (yielding 150 unique quotes) and grouped them based on their capabilities using the affinity diagramming method. Two researchers collected and analyzed data on 51 bots, categorizing them based on their functionalities and publicly available descriptions or marketplace listings.

\newtext{First, data from a randomly selected subset of 10 out of 51 bots was set aside to test the saturation of the hierarchy.}
The researchers collaboratively analyzed the functionalities of the remaining bots, grouping them into higher-level themes through an iterative process. This process was repeated across two additional rounds, refining the themes each time. By the third and final round, the researchers concluded that no new high-level themes could emerge, indicating thematic saturation in the affinity diagramming process. The final hierarchy consists of two levels, as shown in Table~\ref{bots_affinity_diagram}. At the highest level (Level 1), the bots were classified into six overarching themes. Dependency Management, PR Management, CI/CD, Security, Utility, and Miscellaneous. The utility category was further refined in Level 2, where its subclasses elaborated on the themes described in Level 1, offering more precise and specific categorisation of bot functionalities. \newtext{(Details given in Table~\ref{Bots_classes_description}, Appendix \ref{app_rq2})}

The hierarchy of themes provides a comprehensive view of bot functionalities, ranging from automating dependency updates and managing pull requests to addressing security concerns and ensuring smooth integration and delivery workflows. Interestingly, despite analyzing 51 bots across various functionalities, we observed that only four bots are specifically involved in addressing security-related purposes. This finding is notable considering the widespread presence of user-reported security-related issues, suggesting that bot involvement in this critical domain remains minimal.

\snewtext{To better understand their capabilities, we analyzed the source code of 40 (out of 51) bots with publicly available codebases. We manually downloaded these repositories from GitHub (covering various tech stacks) and examined their implementation. Our analysis involved inspecting the working principle from source code to determine whether the bots relied on static rules, pattern matching, or learning-based decision making. We also reviewed their dependencies to examine which external libraries they relied on, particularly looking for any use of machine learning frameworks. Through this systematic process, we found that 85\% (34 bots) relied entirely on pre-defined, rule-based logic, offering limited adaptability to diverse and evolving security issues. Six bots incorporated AI/ML in some form---however, one of them was paid, and two were still in beta stages with restricted functionality. This highlights a clear gap between the potential of AI/ML in security automation and its limited real-world adoption among existing bots. Table \ref{bots-rulebased-cls}, Appendix-\ref{app_rq2} summarizes their types and underlying mechanisms.}

\subsubsection{Under-adoption of bots for detecting and mitigating security-related issues}\label{sec_5_3_2}
\noindent
Among the 51 bots analyzed, only four \textit{viz.} RoboCop, mend-for-github-com, github-advanced-security and Socket Security Bot were identified as actively addressing security-related issues. However, the extremely limited availability and adoption are due to two main factors:

First, all these bots are paid services, which may deter smaller teams or open-source projects from using them. Second, these bots rely heavily on static analysis techniques, which, while valuable, may not adequately address dynamic or runtime vulnerabilities. Consequently, the under-utilization of such bots highlights a significant gap in leveraging automated tools for detecting security issues within GitHub repositories of npm packages.

\section{Characterizing the Interaction Between the Users and Maintainers of GitHub Repositories of npm Packages  (RQ3)}\label{sec:rq3}

We have already identified the different phases involved in the resolution of an issue. Now we check how the community interacts with the maintainers of the repositories. 

\subsection{During creation phase}

Our qualitative analysis identified two broad themes while investigating the types of information provided by a user while reporting a security issue. 

\vspace{1mm}
\noindent
\textbf{Interaction for security issues while offering a solution:} This includes the pull requests made by users. Most of them have a proper description of how it solves the problem. In an issue of storybookjs~\cite{Storybookjs}, the user tells clearly about how the problem is solved \textit{``What I did: Bumps mdx2-csf which resolves a security issue in `loader-utils`''}. Moreover, some of them contain instructions regarding testing their code as in an issue of superset~\cite{Apache}\textit{``\#\#\# TESTING INSTRUCTIONS 1) Visit an Embedded Dashboard, 2) Generate a screenshot of the Dashboard 3) The screenshot should be downloaded successfully''}. However, sometimes they do not contain any description.  Only a title is provided in such cases as in case of an issue of grafana~\cite{Grafana} \textit{``Issue title: Move out Server Admin to a separate menu item on SideMenu''}.

\vspace{1mm}  
\noindent
\textbf{Interaction for security issues without a solution:} These include issues that describe the problem that a user had while using that particular library. These are further classified based on their ability to reproduce the problem from their description. Usually, reproducible issues are provided with code snippets related to the concerned issue. In an issue of apollo-client package~\cite{Apollographql} and pulumi-azure-native~\cite{Pulumi}, users have provided the code snippets to reproduce the issue. Furthermore, users often provide proper steps for the reproduction of the issue. For example, in wp-calypso~\cite{wp-calypso}, the initiator of the issue mentions the steps of reproduction as \textit{``1) Starting at URL: https://wordpress.com/me/security/connected-applications, 2) Click on any of connected application 3) See list of Access Permissions, 4) Switch to another interface language and verify step 3 again''}. Apart from that, users often tend to report the error logs, as reported in an issue of expo package~\cite{Expo} \textit{
``Unhandled promise rejection: Error: Could not encrypt/decrypt the value for SecureStore] at node\_modules/reactnative/Libraries/BatchedBridge/ NativeModules.js:106:50 in promiseMethodWrapper ... ''}. Users even describe their system information in which they have faced the issues as in case of an issue reported in strapi package~\cite{Strapi} \textit{``\#\#\# System Node.js version: 16.13.2, NPM version: 8.1.2, Yarn version: 1.22.15, Strapi version: 4.0.5, Database: PostgreSQL, Operating system: Alpine''}.

However, all issues reported are not provided with proper reproduction steps. Most of them contain some form of feature requests. For example, in their package~\cite{Strapi_2}, a user reports a feature request as \textit{``Is Theia interested in expanding their ESLint config to include XSS sink scanning. 
 Feature Description: I work on Cloud Shell (ide.cloud.google.com).  We use Theia to build our editor.  We have created an eslint config that has generated a list of XSS sinks in Theia (which I am currently sending fixes out for). Are you all potentially interested in having this upstreamed?''}. 

\noindent

\subsection{During discussion phase}

In the qualitative analysis of the discussion phase, repository maintainers responded to the issue in one of two broad themes: they either acknowledged it or ignored it. 

\vspace{1mm}
\noindent
\textbf{Acknowledged the security issue:} The maintainers acknowledge an issue either positively or negatively. Positive acknowledgments involve discussing the problem, such as recognizing they faced the same issue, accepting it, reproducing it, or requesting more details about how the attack works. In an issue of storybooks~\cite{Storybookjs_2}, a user says he has also faced the same issue \textit{``... I'm facing the same issue in `npm audit'. This is in version `5.3.18' of `\@storybook/addon-info'...''}. They may also discuss solutions by resolving the issue in the discussion, suggesting changes, asking for corrections, or agreeing with the proposed solution. For example in an issue of pulumi~\cite{Pulumi_2}, maintainer mentioned \textit{``This is likely due to the way the system treats `id' as a special property... I would suggest using the `RandomString' resource (and then using the language-specific libraries to generate whatever appropriate output format you want in terms of using the value as base64 or hex or whatever else you like)''} 

Negative acknowledgements include marking the issue as a duplicate. They may also involve disagreeing with the suggested solution, rejecting the issue as a false positive, or deeming it not relevant. A comment to saying about the duplicity of the issue in agoric~\cite{Agoric}, \textit{``dup of \#6007''}. Additionally, maintainers may request adherence to policies regarding the disclosure of security vulnerabilities. Like in an issue of aspnetcore~\cite{Dotnet}, \textit{``Thanks for contacting us. While this may be a great idea, it is not aligned with our long-term vision to make it part of the framework. For many other ideas which don't belong to the framework we encourage the community to build and ship on their own, contributing to the expanding .NET Ecosystem...''}\newtext{--- while this reflects a scope decision, it may feel dismissive to the user as no alternative is suggested.}

\vspace{1mm}
\noindent
\textbf{Ignored the security issue:} Maintainers may ignore the issue by showing a lack of interest, which includes not encouraging efforts to solve the problem or providing no comments or engagement. For example, In spartacus~\cite{Sap}, a maintainer commented \textit{``Dropping bug label as this is not a Spartacus bug, nor something we can fix in Spartacus. It's a backend bug. It's also a niche, and we shouldn't spend time here imho.''} \newtext{The wording may sound dismissive due to lack of guidance or follow-up.}

\subsection{During resolution phase}
During the resolution phase, repository maintainers typically close issues under two broad categories: with a valid reason or without one.

\vspace{1mm}
\noindent
\textbf{Closed with valid reason: } Issues may be closed for a variety of reasons, such as false creation, successful resolution, or deferred completion. Falsely created issues include those marked as false positives or those accidentally created without verifying if they had been previously reported. For example, in SonarJS~\cite{SonarSource}, a maintainer commented before closing the issue: \textit{``It seems this was already taken care of by https://github.com/SonarSource/SonarJS/pull/4480''}. A similar, example was seen in RocketChat~\cite{RocketChat}, where a maintainer commented \textit{``Looks a duplicate of \#3069''}. Successfully resolved issues are closed after resolving the problem in discussions, merging a pull request (PR), or referencing a related PR or commit made after the issue was reported. For example, in MetaMask snaps~\cite{MetaMask}, a maintainer resolved the issue in discussion saying \textit{``Consider using proxies instead of hardening each individual endowment with custom wrappers, and only harden the return if it returns something other than itself''}. Another instance where an issue is closed involves merging a pull request is found in netlify~\cite{Netlify}. Some issues remain open for future action, categorized as "to be completed," which may include deferring the fix to a later version or deciding not to address the issue due to its complexity or infeasibility. An issue in joystream~\cite{joystream} is closed by mentioning \textit{``Nothing severe''}. Another example found in npm~\cite{npm2} \textit{``However! It's also not incredibly high urgency for the team. There is a security dimension to this (in that bad actors could shadow important system binaries for nefarious purposes), but that falls into the category of "high impact, low probability" threats...''}.

\vspace{1mm}
\noindent
\textbf{Closed without a valid reason: } Issues are sometimes closed simply because of staleness. E.g., an issue of Dotnet~\cite{Dotnet_2}, marked as stale with the last comment before closing \textit{`` This issue has been automatically marked as stale because it has been marked as requiring author feedback but has not had any activity for 4 days. ''}. \newtext{Although the closure followed an automated policy, doing so after just 4 days of inactivity may feel abrupt to users.} In some cases, issues are closed even without any comment in issues of flow-core~\cite{Ollionorg}, Grafana~\cite{Grafana_2}.

\section{Factors correlating the resolution of security related issues (RQ4)}\label{sec:rq4}

\noindent
Security-related issues often experience extended resolution times, as evidenced by the slower closure of the 13,835 security-tagged issues. Additionally, we observed that issues with CVE tags tend to have faster resolution. Thus, we naturally asked the question--What are the different factors that impact the resolution of a security issue? 

To that end, we developed three Generalized Linear Mixed Models(GLMMs), each targeting one of the three dependent variables--time to close, staleness, and successful resolution. The independent parameters included: (i) CVE mentions, (ii) number of comments (iii) weekly downloads, (iv) number of active maintainers in the past year, (v) issue reproducibility, and (vi) bot involvement. A detailed description of these factors is presented in Table~\ref{tab:factor_desc} of Appendix~\ref{app_RQ4}. Repository-level variations were modelled as a random effect. Results for the models are summarized in Tables \ref{results_glmm_1}, \ref{results_glmm_2}, and \ref{results_glmm_3} of Appendix \ref{app_RQ4}. Next, we present the key takeaways. 

\subsection{Presence of \newtext{CVE} reduces time to close } 

The presence of a \newtext{CVE}
in an issue body significantly decreases the time to close. This prioritization likely stems from their critical nature and potential security implications. 
Issues mentioning a \newtext{CVE}
had a mean time to close of 174.08 days and a median of 70 days, compared to a mean of 307.19 days and a median of 127 days for issues without such mentions. \removetext{\textcolor{red}{\st{Furthermore, 96.26\% of issues with such mentions were not stale, with only 322 out of 8,614 such issues becoming stale.}}} These findings highlight the role of \newtext{CVE}
in expediting issue resolution and reducing the risk of stagnation.

\subsection{Reproducibility prolongs closure and increases staleness, and hinders successful resolution} 
\textbf{Reproducibility increases time to close:} Reproducibility has a significant impact on the time to close, staleness, and successful resolution of issues. Reproducible issues tend to have higher time to close because they require additional time for proper verification and reproduction of the problem.

\snewtext{While reproducibility is typically considered helpful for resolution of issues~\cite{rocklinMinimalBugReports, labelMRE}, our analysis reveals a counterintuitive finding: reproducible issues often remain open longer.} For instance, a long-standing issue in angular.js~\cite{Angular_2} opened in 2020 remained unresolved for years and was only closed in 2024 by a bot due to inactivity. The bot commented: \textit{``This issue has been automatically locked due to inactivity. Please file a new issue if you are encountering a similar or related problem. Read more about our automatic conversation locking policy''}. Another case, found in expressjs~\cite{expressjs} involved a maintainer commenting, \textit{``Yea, this pull request has just lagged around here for years, and seems like we're not going to merge it at this point, and so no reason to keep a stale pull request hanging around for no reason :)''}, which indicates how long delays may occur in issues being abandoned, even if they are reproducible. In terms of stats, reproducible issues have a mean closing time of 105.16 days (median 8 days), while non-reproducible issues close in a mean of 47.21 days (median 2 days), highlighting that reproducibility delays resolution.

\vspace{1mm}

\noindent\textbf{Reproducibility increases staleness:} \snewtext{Contrary to expectation, reproducibility does not always expedite resolution. Instead,}  reproducible issues are also more likely to become stale, especially when there is a lack of capacity to reproduce the issue or when the issue is not addressed promptly. For example, in angular.js~\cite{Angular}, a maintainer closed the issue after seven days of no feedback, stating: \textit{``I'm going to close this issue because we haven't got any feedback. Leave a comment if you can provide new feedback''}. This shows that when an issue cannot be reproduced or lacks updates, it is more likely to be marked as stale. Similarly, other issues have been closed due to the maintenance of repositories when the maintainers no longer considered them relevant or reproducible. One such case is found in storybookjs~\cite{Storybookjs_3}, where a comment stated, \textit{``We’re cleaning house! Storybook has changed a lot since this issue was created and we don’t know if it’s still valid. Please open a new issue referencing this one if this is still a problem in SB 7.x / you can provide a consistent reproduction in 7.x''}.

\vspace{1mm}

\noindent\textbf{Reproducibility hinders successful resolution:} Reproducibility can also decrease the likelihood of a successful resolution. Issues that remain unresolved for a long time due to reproducibility delays tend to be marked stale. In terms of stats, 16,543 reproducible issues were found to be marked as stale. On the other hand, such issues were also closed with an assurance of taking care of them in the next release. For example, a case found in pact-js~\cite{pact-foundation} where a maintainer closed the issue saying \textit{``Thanks for your help on this one. \#282 is being released as we speak (v8.0.5), closing''}. Moreover, many of the reproducible issues were closed with limited conversation, due to prolonged inactivity or because they were already addressed in a newer release. 

\vspace{1mm}

\noindent \textbf{Why reproducibility in issues might not help resolution?} \textcolor{black}{We found that reproducible issues with \newtext{CVE}
references show faster resolution times---mean time to close for reproducible issues with \newtext{CVE} 
mention is 70.75 days (median: 5 days), whereas those without \newtext{CVE}
mention have a mean time to close of 105.6 days (median: 8 days). \snewtext{This further highlights that the benefit of reproducibility is contextual—it helps when developers are sufficiently motivated or equipped to act promptly.} Thus even reproducible security-issues, without \newtext{CVE}
may face challenges in achieving timely resolutions, often due to external constraints or a lack of response from the developers.}

\subsection{\newtext{Bot involvement is correlated with increased staleness and reduced resolution}}
\snewtext{The involvement of bots in Github discussions is strongly correlated with increased staleness and lower chances of successful resolution. Bot activity appears in 72.37\% of stale issues (26,945/37,230) and 37.27\% of unresolved issues (287,028/770,014).} Bots typically intervene during the later stages of an issue's lifecycle, often marking it as stale or automatically closing it, thereby limiting further resolution efforts. For example, in an issue of opensea-js~\cite{ProjectOpenSea}, a bot commented \textit{``This issue has been automatically marked as stale because it has not had recent activity. It will be closed in 14 days if no further activity occurs. Thank you for your contributions. If you believe this was a mistake, please comment``}. \snewtext{This demonstrates how bot involvement aligns with the final stages of an issue. It is potentially associated with  staleness and automated closure after periods of inactivity.}

\subsection{No impact of active maintainers}
Our analysis indicates that the resolution of an issue is independent of the number of active maintainers. This could be attributed to the limited participation of active maintainers in discussions. Moreover, we found the number of active maintainers commenting on an issue to be very low for most of the repositories. Specifically, the mean and median numbers of active maintainers are 4.83 and 2, respectively, while those actively engaged in commenting are only 3.78 and 1. To further strengthen our claim, we computed Pearson's product-moment correlation coefficient between them and found it to be statistically significant ($R = 0.9918$, $p = 0.0$). This confirms our claim that very few active maintainers are engaged in the process of resolution of an issue. 

\section{Implications}

Our study of identifying security issues over npm packages unearths a variety of important observations. To that end, we point out the implications of our study for npm package owners, maintainers, and the open-source community in general.

\vspace{1mm}
\noindent
\textbf{Tagging of issues in GitHub is not uniform: }We know that tagging of issues in GitHub is a fairly common phenomenon \newtext{~\cite{github_labels,lunny_sane_labels}} but is underutilized. Surprisingly, all repositories of our npm dataset do not use tags. Out of 37,278 unique repositories, only 13,031 utilized one or more tags. We found that out
of 10,907,467 issues, around 50\% (5,454,149 issues) does not
contain any tags. As a result of this, we found that only 0. 13\% of 10M+ issues were tagged as \textit{security}. This observation highlights the importance of implementing a standardized tagging system across repositories. It is crucial for repository maintainers to consistently and accurately apply these tags to effectively categorize issues.

\vspace{1mm}
\noindent
\textbf{Bots are inefficient in resolving security-related issues: }Bots, while actively involved in creating security-related issues, typically operate based on predefined rule-based systems. Despite its limitations, these bots are also involved in addressing user-reported security issues. 
\newtext{However, only a limited number of bots claim to support such security concerns, and most of these are paid solutions with limited effectiveness.}
\snewtext{Our analysis further reveals limited adoption of AI/ML in these bots (for both general and security-related functionalities), which stands in contrast to prior work emphasizing the promise of AI/ML techniques for improving security testing~\cite{10.1145/3475716.3475781,10.1007/978-3-031-70879-4_14}.} This underscores the need for the development of advanced and efficient bots to handle security challenges.

\vspace{1mm}
\noindent
\textbf{Reproducibility, CVE mention and bot involvement influence resolution of security-related issues:}
Security-related issues often experience extended resolution times due to additional time required for verifying and reproducing the issues. Moreover, reproducible issues are more likely to become stale because of a lack of capacity to reproduce or feedback. Thus, such issues are not being successfully resolved. On the other hand, issues with CVE mention prompt faster resolution due to its critical nature and potential security implications. Bots, however, are found to negatively impact the resolution process as they often mark the issues as stale or close them which limits further resolution efforts.

\section{Recommendations for Stakeholders}
Finally, we present a set of recommendations tailored for key stakeholders in the context of npm package repositories, including package owners and maintainers, developers of bots for the npm ecosystem, security researchers and end users.

\vspace{1mm}
\noindent
\textbf{Recommendation for npm package owners and maintainers: }
Our findings highlight that the tagging of issues in GitHub repositories is inconsistent and often underutilized. To improve this, package maintainers should implement a standardized tagging system for better categorization, especially for critical issues like security vulnerabilities. \newtext{For maintainers, we recommend involving more active contributors and prioritizing CVE-tagged issues to ensure timely attention to known vulnerabilities.} Automated tools can aid with consistency, while active involvement with contributors and the security research community can address the gaps. Clear tagging guidelines and regular feedback from users will enhance collaboration and enhance the security and functionality of GitHub repositories of npm packages.

\vspace{1mm}
\noindent
\textbf{Recommendation for bot developers for npm ecosystem: } 
\newtext{Our findings highlight that current bots, while involved in creating and addressing security-related issues, often operate on limited, rule-based systems that lack the flexibility to handle complex security concerns. Bot developers should improve their systems using smarter technologies involving machine learning or AI. We also recommend developing bot-based recommender systems (e.g., usable CodeQL~\cite{codeql} tools) to assist users and maintainers during issue reporting—such tools could identify known CVEs or similar issues from other repositories, helping reduce duplication and streamline triage. Moreover, bots must enhance message clarity by including suitable tags, informative descriptions, and reproducible steps.}
Furthermore, given that the entire npm ecosystem is open source, bots should be developed as free solutions, ensuring accessibility to all users while maintaining high-quality, cost-effective support for addressing security challenges.

\vspace{1mm}
\noindent
\textbf{Recommendations for the security research community:} Our results show that critical security issues, despite being reported in the CVE database, are often not properly tagged in multiple issues, leading to their negligence \newtext{(see Section-\ref{sec:rq1})}. This oversight can have serious repercussions, as many packages are directly affected, along with all other packages depending on them. Therefore, better management and prioritization of security issues are needed to ensure that vulnerabilities are promptly addressed and their broader impact is fully recognized. 
\newtext{ To support this, the security research community could develop tools not only for identifying and prioritizing overlooked issues but also for automating issue reproduction, verification, and early detection of false positives—thereby improving the accuracy and efficiency of vulnerability triage for both users and maintainers.}

\vspace{1mm}
\noindent
\textbf{Recommendation for users:} We have already highlighted that reproducibility hinders the resolution process. To that end, the users who are raising security-related issues should provide a proper step-by-step guide towards reproducing the problem, which will aid the maintainers in understanding and verifying the issue quickly. Moreover, they should be more prompt towards replying to queries of maintainers. This will reduce the chances of such issues being marked as stale. On the other hand, users should do proper research before raising issues, as we have identified numerous cases where the issues are marked as false positives or duplicates of existing issues. 

\section{Conclusion}

This study investigated the inefficiencies in addressing security-related issues within the npm ecosystem by analyzing 10,907,467 issues across 45,466 npm packages. Our findings revealed a stark gap in tagging practices, with only 0.13\% of issues explicitly labeled as security-related, despite manual and machine-learning-based analyses identifying 14.8\% as relevant to security. This difference shows that we urgently need a common way to tag issues to improve how we organise and solve them. Furthermore, we identified the constraints of automated systems, which frequently depend on inflexible, rule-based frameworks. This reliance can result in the misclassification of issues as obsolete or the premature closure of cases, thereby hindering efforts towards effective resolution. Reproducibility problems slowed down fixing issues, while CVE comments helped solve problems faster because they have a set level of importance.

These insights call for smarter tools, standardized tagging, and collaborative efforts among stakeholders to enhance security management. Addressing these gaps would facilitate a safer and secure npm ecosystem that more effectively serves both developers and users.

\section{Ethical Considerations} 

\newtext{We initially consulted the ethics committee of our institution regarding our study protocol. They indicated that ethical review was not required, as we collected and analyzed data from publicly accessible npm and GitHub repositories. However, we recognize that this exemption does not, by itself, ensure ethical research. Collecting and analysing public data can still raise ethical concerns, especially when it was not created with research purposes in mind ~\cite{BUCK2021102655}. To that end, we made every effort to conduct our research responsibly---minimizing potential harm and maximizing potential benefits for stakeholders such as npm repository developers, users who contributed content in our dataset, and the platforms (e.g., GitHub and npm) themselves.}

\newtext{In line with prior work on software repository analysis ~\cite{10.1145/3530019.3530041,9978190}, we relied exclusively on data publicly shared by users in \textit{open forums} (forums where no registration was required to access content, e.g., the npm repository and comments on their corresponding GitHub repositories) such as GitHub issues, pull requests, and discussions. We strictly adhered to the ethical principles outlined in previous studies ~\cite{Eysenbach1103}, avoiding any private communications or data collection behind authentication walls. Specifically, all data was obtained via GitHub and npm’s official APIs, and we strictly followed their rate limits and terms of service.}

\newtext{Although the use of public data is common in research ~\cite{BUCK2021102655}, ethical concerns can still arise---particularly when users have not explicitly consented to having their contributions analyzed. To mitigate such risks, we took several precautions. Data was securely stored on password-protected machines physically located within our institution's firewall, with access limited to only our research team. We did not perform any analysis targeting specific developers or repositories. }

\newtext{Our dataset contains the reporting, discussion, and debate of issues specifically within open-source projects, often conducted in public forums like GitHub. These disclosures reflect broader open-source community norms around transparency and shared responsibility. One of our primary goals is to support and align with these norms by facilitating a more informed and nuanced conversation about the ethical use of such data to uncover high-level data-driven insights which can benefit the open source projects.}

\newtext{Thus, we strongly believe our research has been conducted ethically and contributes valuable insights. We hope that these findings can help strengthen collaboration and security practices within the open-source ecosystem---highlighting the potential benefits of this work to the broader community.}

\section{Compliance with Open Science Policy} 

We will adhere to open science policy. We are releasing the artifacts (models) with accompanying code so that the results of this work can be reproduced and the models can be used in further research (e.g., for discovering security-related issues). Since the raw data might contain identifying information about security issues, to address potential ethical concerns about open science compliance, we are sharing the raw datasetsxs only with verified researchers upon request, rather than releasing them publicly. We are providing clear instructions about accessing the raw data for academic research to researchers on providing informed consent about acceptable data usage. For accessing the models and instructions to access data, please visit \url{https://doi.org/10.5281/zenodo.15614029}

\section{Acknowledgement}

We thank the anonymous reviewers and our shepherd for their valuable feedback. We also thank Gunjan Balde for his help throughout this work. The authors thank everyone who facilitated access to the necessary resources during the data collection from GitHub. This research was (partially) funded by a Google India Faculty Research Award.

\bibliographystyle{plain}
\bibliography{main}

\appendix
\section{Details of npm packages}\label{app_sec2} 
\begin{figure}[h]
    \scriptsize
    \centering
    \includegraphics[width=\linewidth]
    {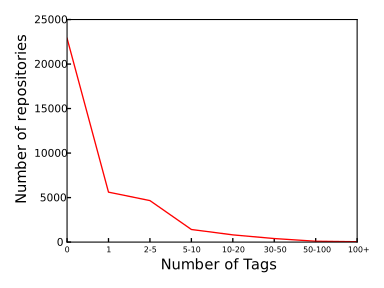}
    \caption{Distribution of repositories across different tag count ranges.}
    \label{freq_tag_repo}
\end{figure}

In this section, we provide additional details related to the dataset used in our study, including examples of popular npm packages from different buckets, as shown in Table \ref{pop_bucket}. Furthermore, we present the set of tags identified as similar to \textit{security} in Table \ref{tags_sec_w2vec}, along with the distribution of tags per repository depicted in Figure \ref{freq_tag_repo}.

\begin{table}[h]
\scriptsize
\centering
\begin{tabular}{|c|l|}
\hline
\rowcolor[HTML]{C0C0C0} 
\textbf{Indegree of packages}        & \textbf{Popular Javascript packages}      \\ \hline
                                     & @ever-co/faker                            \\ \cline{2-2} 
\multirow{-2}{*}{0}                  & @musicstory/react-bootstrap-table2-filter \\ \hline
                                     & @aurelia/scheduler-dom                    \\ \cline{2-2} 
\multirow{-2}{*}{1-10}               & @tanstack/svelte-table                    \\ \hline
                                     & @ckeditor/ckeditor5-vue                   \\ \cline{2-2} 
\multirow{-2}{*}{10-100}             & create-vite                               \\ \hline
                                     & @medusajs/medusa                          \\ \cline{2-2} 
\multirow{-2}{*}{100-500}            & gatsby-plugin-manifest                    \\ \hline
                                     & @types/chalk                              \\ \cline{2-2} 
\multirow{-2}{*}{500-1000}           & markdown                                  \\ \hline
                                     & ethers                                    \\ \cline{2-2} 
\multirow{-2}{*}{\textgreater{}1000} & shelljs                                   \\ \hline
\end{tabular}
\caption{Popular packages in different buckets.}
\label{pop_bucket}
\end{table}

\begin{table}[]
\centering
\scriptsize
\begin{tabular}{|l|l|}
\hline
\rowcolor[HTML]{C0C0C0} 
\textbf{Issue age} & \textbf{\begin{tabular}[l]{@{}l@{}}\# issues \\ created \end{tabular}} \\ \hline

3 months                   & 35284                                                                               \\ \hline
6 months                   & 60082                                                                              \\ \hline
9 months                   & 60692                                                                              \\ \hline
1 year                     & 71675                                                                              \\ \hline
2 years                    & 253022                                                                             \\ \hline
3 years                    & 253703                                                                             \\ \hline
> 3 years                  & 883280                                                                            \\ \hline
\end{tabular}
\caption{Distribution of issues based on their creation time.}
\label{table:Issues_created_back_ago}
\end{table}

\begin{table}[h]
\scriptsize
\centering
\begin{tabular}{|l|}
\hline
\rowcolor[HTML]{C0C0C0} 
\textbf{Tags} \\ \hline
category: security \\ \hline
type: security \\ \hline
security \\ \hline
security (category) \\ \hline
theme: security \\ \hline
severity6: security \\ \hline
:label: security \\ \hline
pr: security \\ \hline
:fire: security \\ \hline
area: security \\ \hline
issue: security \\ \hline
{[}type{]} security \\ \hline
security :exclamation: \\ \hline
topic: security \\ \hline
cat: security \\ \hline
severity: security \\ \hline
subj: security \\ \hline
security vulnerability \\ \hline
3: security review \\ \hline
security fix \\ \hline
impact/security \\ \hline
limitation: security restriction \\ \hline
mend: dependency security vulnerability \\ \hline
security blocker \\ \hline
external: lightning web security \/ locker \\ \hline
\end{tabular}

\caption{Set of tags that are found similar with the term \textit{security} using Word2vec.}
\label{tags_sec_w2vec}
\end{table}

\section{Security-issues identification model}~\label{sec-issu}
\textbf{Models considered to detect security-related issues:} To classify issues into two classes (related to security or not related to security), we used the following models: BERT~\cite{devlin2018bert}, RoBERTa~\cite{liu2019roberta}, CodeBERT~\cite{feng2020codebert}, FLAN-T5~\cite{chung2022scalinginstructionfinetunedlanguagemodels} and DeBERTa~\cite{he2020deberta} and LLMs namely Mistral~\cite{jiang2023mistral}, Qwen~\cite{qwen}, Meta-LLaMA~\cite{touvron2023llama}, and Gemma~\cite{team2024gemma} using zero-shot and few-shot under ICL settings. Since the task is a sequence classification task and hence, we used different architectures of models described in Table \ref{details_bertmodels_arch} along with the specification of model sizes. The hyperparameter tuning was done using Optuna ~\cite{akiba2019optunanextgenerationhyperparameteroptimization},
which is a framework for optimization for hyperparameters.
The performance of the models is shown in Table \ref{Performance_finetuned_bertmodels}.

\begin{figure}
\scriptsize
    \centering
    \includegraphics[width=1\linewidth]{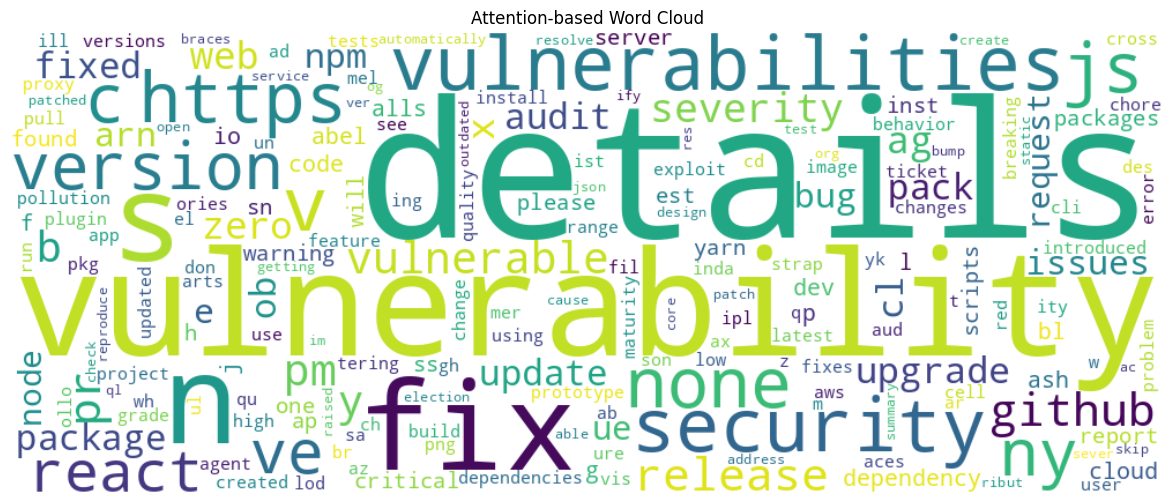}
    \caption{Wordcloud of most influential tokens identified using [CLS] token attention scores from the final layer of our fine-tuned RoBERTa model.}
    \label{fig:wordcloud}
\end{figure}

\snewtext{We extracted and averaged attention scores from the [CLS] token across all heads in the final layer of the fine-tuned RoBERTa model to identify influential tokens, which are visualized in the word cloud below (see Figure-\ref{fig:wordcloud}) after removing stopwords and artifacts.}

\noindent
\textbf{Error Analysis of RoBERTa}: Since we achieved the best accuracy for the downstream task using the fine-tuned RoBERTa model. Class-wise performance of the RoBERTa model is shown in Table \ref{Roberta_classwise_performance}. We also tried to get insights into the performance of the model using error analysis and observed that the issue body often lacks sufficient information, with references to other issues by their numbers instead of detailed elaboration. Additionally, while discussions occasionally address relevant topics, they often lack the expected focus on security-related context. There is also a noticeable drift between issue titles, which focus on security, and the corresponding issue bodies, which do not align with this context, further complicating accurate classification.

\noindent
\textbf{Performance of LLMs: } For LLM evaluation, task-specific prompts were crafted to address the classification task. Zero-shot settings relied on these prompts without additional fine-tuning, while few-shot ICL incorporated a small number of annotated examples into the prompts to provide better context. The prompts for zero-shot and few-shot evaluations are detailed in Table \ref{llm_prompts} and the specific LLM models evaluated are described in Table \ref{llm_details}. Model outputs were assessed using a BERT-based similarity scoring approach~\cite{zhang2019bertscore} with the DeBERTa~\cite{he2020deberta} model, comparing predictions against reference sentences, \textit{security} and \textit{non-security}. The evaluation results are given in Table \ref{Performance_finetuned_llms}.

\begin{table}[h]
\scriptsize
\centering
\begin{tabular}{|c|c|c|c|}
\hline
\rowcolor[HTML]{C0C0C0} 
\textbf{Model}            &           & \textbf{Accuracy} & \textbf{Macro avg F1} \\ \hline
                          & Zero Shot & 0.82              & 0.82                  \\ \cline{2-4} 
\multirow{-2}{*}{Mistral} & Few Shot  & 0.83              & 0.83                  \\ \hline
                          & Zero Shot & 0.52              & 0.39                  \\ \cline{2-4} 
\multirow{-2}{*}{Qwen}    & Few Shot  & 0.67              & 0.66                  \\ \hline
                          & Zero Shot & 0.65              & 0.63                  \\ \cline{2-4} 
\multirow{-2}{*}{LLaMA}   & Few Shot  & 0.78              & 0.77                  \\ \hline
                          & Zero Shot & 0.60               & 0.53                  \\ \cline{2-4} 
\multirow{-2}{*}{Gemma}   & Few Shot  & 0.54              & 0.50                   \\ \hline
\end{tabular}
\caption{Performance of large language models under zero-shot and few-shot ICL settings for classifying issues into security-related and non-security-related categories.}
\label{Performance_finetuned_llms}
\end{table}

\begin{table}[h]
\centering
\scriptsize
\begin{tabular}{|
>{\columncolor[HTML]{C0C0C0}}l |l|l|}
\hline
\textbf{Model}   & \cellcolor[HTML]{C0C0C0}\textbf{Model Specification} & \cellcolor[HTML]{C0C0C0}\textbf{Task Type} \\ \hline
\textbf{Gemma}   & google/gemma-2b-it                                   & Text Classification                        \\ \hline
\textbf{Mistral} & mistralai/Mistral-7B-Instruct-v0.3                   & Text Classification                        \\ \hline
\textbf{Qwen}    & Qwen/Qwen2.5-1.5B-Instruct                           & Text Classification                        \\ \hline
\textbf{Llama}   & meta-llama/Llama-3.2-3B-Instruct                     & Text Classification                        \\ \hline
\end{tabular}

\caption{Specification of the LLMs used for classification of security-related and non-security related issues under zero-shot and few-shot ICL settings.}
\label{llm_details}
\end{table}

\begin{table}[h]
\scriptsize
\centering
\begin{tabular}{llll}
\cline{3-4}        & \multicolumn{1}{l|}{}             & \multicolumn{2}{c|}{\textbf{Predicted}}                                                         \\ \cline{3-4}   & \multicolumn{1}{l|}{}             & \multicolumn{1}{l|}{security}                  & \multicolumn{1}{l|}{non security}              \\ \hline
\multicolumn{1}{|c|}{}                                                                                  & \multicolumn{1}{l|}{security}     & \multicolumn{1}{l|}{4}                         & \multicolumn{1}{l|}{\cellcolor[HTML]{C0C0C0}3} \\ \cline{2-4} 
\multicolumn{1}{|c|}{\multirow{-2}{*}{\textbf{\begin{tabular}[c]{@{}c@{}}Ground\\ Truth\end{tabular}}}} & \multicolumn{1}{l|}{non security} & \multicolumn{1}{l|}{\cellcolor[HTML]{C0C0C0}4} & \multicolumn{1}{l|}{89}                        \\ \hline&                                   &                                                &                                                \\ &                                   &                                                &                                               
\end{tabular}
\caption{Confusion matrix for our DeBERTa text classifier.}
\label{misclassification_Debarta}
\end{table}

\begin{table*}[h]
\scriptsize
\centering
\begin{tabular}{|
>{\columncolor[HTML]{C0C0C0}}l |l|l|}
\hline
\multicolumn{1}{|c|}{\cellcolor[HTML]{C0C0C0}\textbf{Setting}} & \cellcolor[HTML]{C0C0C0}\textbf{Prompt Template}                                                                                                                                                                                                                                                                                                                                                                                                                                                                                                                                                                                                                                                    & \cellcolor[HTML]{C0C0C0}\textbf{Description}                                                                                                                                                                                                                \\ \hline
\textbf{Zero-shot}                                             & \begin{tabular}[c]{@{}l@{}}\{\\        "role": "user",\\         "content": f"classify **WITHOUT EXPLANATION** whether \\          the given issue is related to security or not. Issue: \{new\_issue\}"\\ \}\end{tabular}                                                                                                                                                                                                                                                                                                                                                                                                                                                                          & \begin{tabular}[c]{@{}l@{}}Direct prompt asking the model\\  to classify the issue without \\ providing any explanation.\end{tabular}                                                                                                                       \\ \hline
\textbf{Few Shot}                                              & \begin{tabular}[c]{@{}l@{}}\{\\       "role":"user", "content": "classify **WITHOUT EXPLANATION** whether \\        the given issue is related to security or not. Issue: \{issue\_0\}" \\ \}, \\ \{ \\       "role":"assistant", "content": "security" \\ \}, \\ \{ \\       "role":"user", "content": "classify **WITHOUT EXPLANATION** whether \\        the given issue is related to security or not. Issue: \{issue\_1\}" \\ \}, \\ \{ \\       "role":"assistant", "content": "non security" \},\\  ... \\ \{ \\       "role":"user", "content": "classify **WITHOUT EXPLANATION** whether \\        the given issue is related to security or not. Issue: \{new\_issue\}"\\ \}\end{tabular} & \begin{tabular}[c]{@{}l@{}}Few-shot prompt providing \\ two examples each for "security" \\ and "non-security" classifications \\ alternatively to give the model \\ contextual guidance before asking \\ it to predict for the current issue.\end{tabular} \\ \hline
\end{tabular}
\caption{Prompts for zero-shot and few-shot under ICL settings for classifying issues into security-related and non security-related categories.}
\label{llm_prompts}
\end{table*}

\begin{table*}[h]
\scriptsize
\resizebox{\textwidth}{!}{\begin{tabular}{|ll|ll|}
\hline
\rowcolor[HTML]{C0C0C0} 
\multicolumn{2}{|c|}{\cellcolor[HTML]{C0C0C0}\textbf{Labelled as security}} & \multicolumn{2}{c|}{\cellcolor[HTML]{C0C0C0}\textbf{Model predicted}} \\ \hline
\rowcolor[HTML]{C0C0C0} 
\multicolumn{1}{|l|}{\cellcolor[HTML]{C0C0C0}\textbf{Issue Title}} & \textbf{Issue Body} & \multicolumn{1}{l|}{\cellcolor[HTML]{C0C0C0}\textbf{Issue Title}} & \textbf{Issue Body} \\ \hline
\multicolumn{1}{|l|}{\begin{tabular}[c]{@{}l@{}}Webhook notification\\ channel password field\\ is in plain text\end{tabular}} & \begin{tabular}[c]{@{}l@{}}What would you like to be added: It would be \\ ideal if the password field for the webhook \\ notification channel was hidden and not \\ exposed in plain text. Why is this needed:Anyone \\ who has access to set up notification channels can \\ see the password which raises a security concern \\ for the application we send the alert to.\end{tabular} & \multicolumn{1}{l|}{\begin{tabular}[c]{@{}l@{}}fix: remove unnecessary \\ logging for webauthn \\ auth\end{tabular}} & \begin{tabular}[c]{@{}l@{}}When authenticating using Webauthn \\ it is unnecessary to log Log in on.... \\ This PR hides that logging in that case.\end{tabular} \\ \hline
\multicolumn{1}{|l|}{\begin{tabular}[c]{@{}l@{}}Set Cache-control: \\ no-cache for full \\ page requests\end{tabular}} & \begin{tabular}[c]{@{}l@{}}All API requests already set Cache-control: \\ no-cache to avoid browsers caching sensitive \\ data. The full page requests, however, does\\  not which means that all data in \\ window.grafanaBootData can be cached \\ by the browsers.This means that browsers\\  can cache some data even if the user logged out.\end{tabular} & \multicolumn{1}{l|}{\begin{tabular}[c]{@{}l@{}}Migrate Lambda Trigger is \\ not triggering if \\ USER\_PASSWORD\_AUTH \\ is done on the server/backend...\end{tabular}} & \begin{tabular}[c]{@{}l@{}}Describe the bug If you use the InitiateAuth \\ with the USER\_PASSWORD\_AUTH as the authflow, \\ the Migrate Lambda Trigger will not work... I have a \\ console.log in the lambda and it is not working. \\ However, if I copy-paste the exact code on the browser-side, \\ the trigger will work...\\ SDK version number\\ @aws-sdk/client-cognito-identity-provider@3.294.0\\ Which JavaScript Runtime is this issue in?\\ Node.js\\ Details of the browser/Node.js/ReactNative version\\ v16.19.0\end{tabular} \\ \hline
\multicolumn{1}{|l|}{\begin{tabular}[c]{@{}l@{}}XS deep freeze conflicts\\ with SES security \\ constraints\end{tabular}} & \begin{tabular}[c]{@{}l@{}}The XS Object.freeze takes a second optional \\ boolean parameter that, if truthy, causes some \\ form of transitive freezing. But unlike harden, \\ it does more freezing than user code can do (and \\ inspired thepetrify notion we're still designing). \\ As currently implemented,this enhanced \\ Object.freeze can be used for attack.\end{tabular} & \multicolumn{1}{l|}{\begin{tabular}[c]{@{}l@{}}{[}SECURITY{]} npm i logs bearer\\ token in case there is a \\ formatting issue.\end{tabular}} & \begin{tabular}[c]{@{}l@{}}Is there an existing issue for this?I have searched \\ the existing issues This issue exists in the latest \\ npm versionI am using the latest npm Current \\ BehaviorAccidentally providing a misformed token \\ will print  the bearer token to the log-output.\\ I wasn't sure if this is indeed a security risk but \\ I figured it might not hurt to point it out in case it is. \\ Please close the issue right away if this is not critical.\\ Expected Behavior Do not print any bearer tokens to \\ standard output.\end{tabular} \\ \hline
\multicolumn{1}{|l|}{\begin{tabular}[c]{@{}l@{}}Anonymous users are \\ not redirected to login \\ page after trying to \\ access a checkout step \\ route since 3.0.0\end{tabular}} & \begin{tabular}[c]{@{}l@{}}Old checkout bug behavior since Spartacus \\ version 3.0.0 Checkout auth guard is not \\ redirecting anonymous users that has no \\ active cart id. They get stuck in a blank \\ page if they were to try to visit the checkout \\ route directly\end{tabular} & \multicolumn{1}{l|}{\begin{tabular}[c]{@{}l@{}}Fix for SSID and passwords \\ starting with 0x\end{tabular}} & \begin{tabular}[c]{@{}l@{}}By default string starting with 0x are parsed as \\ hexadecimal numbers by yargs. This fix allows to \\ configure devices with an SSID starting with the string 0x\end{tabular} \\ \hline
\multicolumn{1}{|l|}{\begin{tabular}[c]{@{}l@{}}Disable localhost: snaps \\ outside of Flask\end{tabular}} & \begin{tabular}[c]{@{}l@{}}localhost:snaps are not something that we \\ ever want to support in stable MetaMask, \\ and we should ensure that they are disabled in \\ stable distributions. Outcome: Snaps that are \\ fetched from localhost should only be available \\ in  Flask distribution. The user should be notified \\ that this snap is a  debug one and will not be loaded.\end{tabular} & \multicolumn{1}{l|}{\begin{tabular}[c]{@{}l@{}}fix(lib-storage): S3 Upload can \\ corrupt data from \\ readable stream\end{tabular}} & \begin{tabular}[c]{@{}l@{}}Issue\\ In some circumstance Upload miss some data from \\ the readable stream and corrupt the uploaded data. \\ This has been observed in a real situation when \\ serializing data and uploading data on the fly by \\ using a PassThrough between the serialization and the \\ S3 Upload. The issue has been located in\\  lib/storage/src/data-chunk/readable-helper.ts\\ Description In readable-helper.ts the pause/resume \\ mechanism has a concurrency \\ flaw between line\end{tabular} \\ \hline
\end{tabular}}
\caption{Examples of issues labelled as security and issues that our model predicted to be security related.}
\label{examples_5_5_issues}
\end{table*}

\begin{table}[h]
\scriptsize
\centering
\begin{tabular}{|l|l|l|}
\hline
\rowcolor[HTML]{C0C0C0} 
{\color[HTML]{000000} }                                                                         & {\color[HTML]{000000} \textbf{\begin{tabular}[c]{@{}l@{}}\% themes in\\ validation dataset\end{tabular}}} & {\color[HTML]{000000} \textbf{\begin{tabular}[c]{@{}l@{}}\% themes in additional\\ random sample\end{tabular}}} \\ \hline
{\color[HTML]{000000} \textbf{\begin{tabular}[c]{@{}l@{}}Correctly \\ Predicted\end{tabular}}}  & {\color[HTML]{000000} 82\%}                                                                               & {\color[HTML]{000000} 76\%}\                                                                                     \\ \hline
{\color[HTML]{000000} \textbf{\begin{tabular}[c]{@{}l@{}}Incorrectly\\ predicted\end{tabular}}} & {\color[HTML]{000000} 18\%}                                                                               & {\color[HTML]{000000} 24\%}                                                                                     \\ \hline
\end{tabular}
\caption{Performance of fine-tuned RoBERTa model (For theme classification) on validation dataset and additional random sample (with manual ground truth labels).}
\label{table_themes_roberta_appendix}
\end{table}

\begin{table}[h]
\centering
\scriptsize
\begin{tabular}{|l|l|}
\hline
\rowcolor[HTML]{C0C0C0} 
\textbf{Activities} & \textbf{Frequency} \\ \hline
comment                        & 2826              \\ \hline
labeled                        & 2197              \\ \hline
unlabeled                      & 1079               \\ \hline
milestoned                     & 294                \\ \hline
locked                         & 287                \\ \hline
closed                         & 233               \\ \hline
referenced                     & 182               \\ \hline
review\_requested               & 129               \\ \hline
deployed                       & 63                \\ \hline
demilestoned                   & 45                 \\ \hline
subscribed                     & 19                 \\ \hline
merged                         & 19               \\ \hline
head\_ref\_deleted               & 19               \\ \hline
mentioned                      & 16              \\ \hline
removed\_from\_merge\_queue       & 4                  \\ \hline
head\_ref\_force\_pushed          & 3               \\ \hline
added\_to\_project               & 3                  \\ \hline
moved\_columns\_in\_project       & 2                  \\ \hline
automatic\_base\_change\_succeeded & 2                 \\ \hline
reopened                       & 2               \\ \hline
renamed                        & 2               \\ \hline
review\_dismissed               & 1                 \\ \hline
convert\_to\_draft               & 1                  \\ \hline
review\_request\_removed         & 1                 \\ \hline
assigned                       & 1                  \\ \hline

\end{tabular}
\caption{Overview of the various activities performed by bots in user-reported issues along with the count of their occurrence in security-related issues.}
\label{table:Frequency_of_botactivities}
\end{table}

\begin{table}[h]
\centering
\scriptsize
\begin{tabular}{|c|c|c|l|}
\hline
\rowcolor[HTML]{C0C0C0} 
\textbf{\begin{tabular}[c]{@{}c@{}}Bot Type (L1)\end{tabular}}                   & \textbf{\begin{tabular}[c]{@{}c@{}}Bot Type (L2)\end{tabular} }                & \textbf{\begin{tabular}[c]{@{}c@{}}Bot Type (L3)\end{tabular} }     & \textbf{No of Bots} \\ \hline&                                                                          & \begin{tabular}[c]{@{}c@{}}Public \\ Source Codes\end{tabular}   & 40         \\ \cline{3-4} & \multirow{-2}{*}{\begin{tabular}[c]{@{}c@{}}Public \\ Apps\end{tabular}} & \begin{tabular}[c]{@{}c@{}}Source Code not \\ found\end{tabular} & 17         \\ \cline{2-4} 
\multirow{-3}{*}{\begin{tabular}[c]{@{}c@{}}Hosted as \\ GitHub App\end{tabular}}   & \begin{tabular}[c]{@{}c@{}}Private \\ Apps\end{tabular}                  &                                                               & 14         \\ \hline
\begin{tabular}[c]{@{}c@{}}Hosted as \\ normal GitHub \\ User accounts\end{tabular} &                                                                          &                                                               & 49         \\ \hline
\end{tabular}

\caption{Classification of 120 GitHub bots in user-reported security issues based on their deployment and accessibility.}
\label{tab:bot_classification_usercreated_issues}
\end{table}

\section{Theme identification model} \label{app_theme}

For the identification of themes in reviews, we used the following models: BERT~\cite{devlin2018bert}, RoBERTa~\cite{liu2019roberta}, CodeBERT~\cite{feng2020codebert}, FLAN-T5~\cite{chung2022scalinginstructionfinetunedlanguagemodels} and DeBERTa~\cite{he2020deberta}. Since the task can be modelled as both sequence multi-label classification or sequence generation task, hence, we used different architectures of models described in Table-\ref{details_bertmodels_arch} along with the specification of model sizes. The hyperparameter tuning was done using Optuna ~\cite{akiba2019optunanextgenerationhyperparameteroptimization}, which is a framework for optimization for hyperparameters.

\noindent
\textbf{Accuracy of models}: We found that RoBERTa model outperformed other models into consideration, identifying more than 80\% of themes correctly in the reviews as shown in Table-\ref{model_stats_theme_identification} which reports the accuracy of the models on validation set.The result of the accuracy of RoBERTa model against additional manual ground truth is reported in Table-\ref{table_themes_roberta_appendix} .

\noindent
\textbf{Error Analysis of RoBERTa}: Since the fine-tuned RoBERTa model predicted the highest fraction of themes correctly, we did further analysis to look into the performance of the model. Classwise performance of the RoBERTa model is shown in Table-\ref{classwise_performace_themes}. \newtext{Although recall for “Successfully resolved” is lower, the F1-score (0.68) shows a balanced performance, suggesting accurate predictions; our high-precision model thus provides a conservative lower-bound estimate for this theme.} We also tried to get insights into the performance of the model using error analysis and observed that the model mostly commits mistakes in predicting less frequently occurring classes like speaking against the issue/not interested in the issue. We also noted that the model committed mistakes in understanding the contextual meaning of phrases and made predictions using the phrases themselves.

\begin{table}[h]
\scriptsize
\centering
\begin{tabular}{|l|l|l|l|}
\hline
\rowcolor[HTML]{C0C0C0} 
Theme                      & Hierarchy & F1      & F2  \\ \hline
\rowcolor[HTML]{CBCEFB} 
A. Issue with solution(PR) & L1        & 303 & 826,030   \\ \hline
A.1 Description present    & L2      & 194   & 786,818 \\ \hline
A.2 No description    \textsuperscript{*}     & L2   & 109      & 39,212    \\ \hline
\rowcolor[HTML]{CBCEFB} 
B. Issue without solution  & L1      & 225    & 791,368\\ \hline
B.1 Reproducibility        & L2     & 102   & 418,668  \\ \hline
B.2 Non-reproducibility    & L2      & 123  & 372,700  \\ \hline
\end{tabular}
\caption{The hierarchy of themes derived from open coding of reviews in the category of "creation". Here, F1 refers to the number of quotes in which this theme appeared in our manual analysis and F2 refers to the number of quotes in which the theme was identified by the automated approach.\textsuperscript{*} We filtered the case of Issues with solution but no description logically and did not feed to the model.}
\label{1_5M_inferance_creation}
\end{table}

\begin{table}[h]
\scriptsize
\centering
\begin{tabular}{|l|l|l|l|}
\hline
\rowcolor[HTML]{C0C0C0} 
Theme                        & Hierarchy & F1        & F2  \\ \hline
\rowcolor[HTML]{CBCEFB} 
A. Acknowledged              & L1      & 400  & 1,981,510  \\ \hline
A.1 Spoke against with issue & L2    & 91    & 194,971     \\ \hline
A.2 Spoke for the issue      & L2     & 309   & 1,786,539  \\ \hline
\rowcolor[HTML]{CBCEFB} 
B. Ignored                   & L1      & 150  & 486,081    \\ \hline
B.1 Not interested           & L2       & 46   & 133,035   \\ \hline
B.2 Inconclusive             & L2     & 104    & 353,046   \\ \hline
\end{tabular}
\caption{The hierarchy of themes derived from open coding of reviews in the category of "discussion". Here, F1 refers to the number of quotes in which this theme appeared in our manual analysis and F2 refers to the number of quotes in which the theme was identified by the automated approach. }
\label{1_5M_inferance_discussion}
\end{table}

\begin{table}[h]
\scriptsize
\centering
\begin{tabular}{|l|l|l|l|}
\hline
\rowcolor[HTML]{C0C0C0} 
Theme                          & Hierarchy & F1        & F2  \\ \hline
\rowcolor[HTML]{CBCEFB} 
A. With valid reason           & L1         & 497 & 1,281,922\\ \hline
A.1 Falsely Created            & L2            & 60 & 73,670 \\ \hline
A.2 Successfully resolved      & L2           & 260 & 727,861\\ \hline
A.3 To be completed            & L2           & 177 & 480,391\\ \hline
\rowcolor[HTML]{CBCEFB} 
B. Without valid reason        & L1           & 172  & 116,418\\ \hline
B.1 Closed without reason \textsuperscript{**}   & L2      & 61  & 79,188      \\ \hline
B.2 Completed due to staleness & L2            & 111 & 37,230\\ \hline
\end{tabular}
\caption{The hierarchy of themes derived from open coding of reviews in the category of "resolution". Here, F1 refers to the number of quotes in which this theme appeared in our manual analysis and F2 refers to the number of quotes in which the theme was identified by the automated approach.\textsuperscript{**}We filtered the case of Closed without any reason logically and did not feed to the model. }
\label{1_5M_inferance_resolution}
\end{table}

\begin{table}[h]
\scriptsize
\centering
\begin{tabular}{|l|l|l|l|}
\hline
\rowcolor[HTML]{C0C0C0} 
Type & L2 levels & Recall & F1-score \\ \hline
 & Reproducibility & 0.93 & 0.79 \\ \cline{2-4} 
 & Non reproducibility & 0.79 & 0.87 \\ \cline{2-4} 
\multirow{-3}{*}{Creation} & Description present & 0.97 & 0.93 \\ \hline
 & Spoke for the issue & 0.93 & 0.91 \\ \cline{2-4} 
 & Spoke against with issue & 0.6 & 0.67 \\ \cline{2-4} 
 & Not interested & 0.60 & 0.60 \\ \cline{2-4} 
\multirow{-4}{*}{Discussion} & Inconclusive & 0.81 & 0.87 \\ \hline
 & Completed due to staleness & 1.00 & 0.98 \\ \cline{2-4} 
 & Falsely Created & 0.62 & 0.70 \\ \cline{2-4} 
 & Successfully resolved & 0.53 & 0.68 \\ \cline{2-4} 
\multirow{-4}{*}{Resolution} & To be completed & 0.94 & 0.72 \\ \hline
\end{tabular}
\caption{ Class-wise performance of RoBERTa model on theme identification task on validation dataset.}
\label{classwise_performace_themes}
\end{table}

\begin{table*}[h]
\scriptsize
\centering
\begin{tabular}{|c|c|c|c|}
\hline
\rowcolor[HTML]{C0C0C0} 
{\color[HTML]{000000} \textbf{Model}}    & {\color[HTML]{000000} \textbf{Model Specification}} & {\color[HTML]{000000} \textbf{Classification of issues}} & \textbf{Theme Identification}                                                                          \\ \hline
{\color[HTML]{000000} \textbf{CodeBERT}} & {\color[HTML]{000000} CodeBERT-base}                & {\color[HTML]{000000} Sequence classification}           & \begin{tabular}[c]{@{}c@{}}Sequence Classification\\ (With multiple classification heads)\end{tabular} \\ \hline
{\color[HTML]{000000} \textbf{BERT}}     & {\color[HTML]{000000} BERT-base (Uncased)}          & {\color[HTML]{000000} Sequence classification}           & \begin{tabular}[c]{@{}c@{}}Sequence Classification\\ (With multiple classification heads)\end{tabular} \\ \hline
\textbf{RoBERTa}                         & RoBERTa-base                                        & Sequence classification                                  & \begin{tabular}[c]{@{}c@{}}Sequence Classification\\ (With multiple classification heads)\end{tabular} \\ \hline
\textbf{Flan-T5}                         & Flan-T5-base                                        & Seq2Seq                                                  & Seq2Seq                                                                                                \\ \hline
\textbf{DeBERTa}                         & DeBERTa-v3-base                                     & Sequence classification                                  & \begin{tabular}[c]{@{}c@{}}Sequence Classification\\ (With multiple classification heads)\end{tabular} \\ \hline
\end{tabular}

\caption{Specification and architecture of models used in classification of issues and for theme identification.}
\label{details_bertmodels_arch}
\end{table*}

\section{Temporal Distribution of Issues: Categorization by Creation Time}\label{app_rq1}

\noindent
Distribution of issues based on their creation time, categorizing the number of issues created within specific time frames : \ref{table:Issues_created_back_ago}. Also the five most frequently mentioned CVE IDs (\newtext{strictly CVE-IDs are mentioned here)} are in the Table \ref{cve_5_description}. 

\begin{table*}[]
\scriptsize
\centering
\begin{tabular}{|l|l|l|}
\hline
\rowcolor[HTML]{C0C0C0} 
\textbf{CVE- id} & \textbf{Title} & \textbf{Description} \\ \hline
CVE-2023-45857 & \begin{tabular}[c]{@{}l@{}}Axios Cross-Site Request \\ Forgery Vulnerability\end{tabular} & \begin{tabular}[c]{@{}l@{}}An issue discovered in Axios 0.8.1 through 1.5.1 \\ inadvertently reveals the confidential XSRF-TOKEN \\ stored in cookies by including it in the HTTP header \\ X-XSRF-TOKEN for every request made to any host\\  allowing attackers to view sensitive information.\end{tabular} \\ \hline
CVE-2021-23337 & Command Injection in lodash & \begin{tabular}[c]{@{}l@{}}lodash versions prior to 4.17.21 are vulnerable to \\ Command Injection via the template function.\end{tabular} \\ \hline
CVE-2019-10744 & Prototype Pollution in lodash & \begin{tabular}[c]{@{}l@{}}Versions of lodash before 4.17.12 are vulnerable to \\ Prototype Pollution. The function defaultsDeep allows \\ a malicious user to modify the prototype of Object via \\ \{constructor: \{prototype: \{...\}\}\}causing the addition or \\ modification of an existing property that will exist on \\ all objects.\end{tabular} \\ \hline
CVE-2021-3807 & \begin{tabular}[c]{@{}l@{}}Inefficient Regular Expression \\ Complexity in chalk/ansi-regex\end{tabular} & \begin{tabular}[c]{@{}l@{}}ansi-regex is vulnerable to Inefficient Regular \\Expression Complexity which could lead to a denial \\of service  when parsing invalid ANSI escape codes.\end{tabular} \\ \hline
CVE-2020-28469 & \begin{tabular}[c]{@{}l@{}}glob-parent vulnerable to Regular \\ Expression Denial of Service in \\ enclosure regex\end{tabular} & \begin{tabular}[c]{@{}l@{}}This affects the package glob-parent before 5.1.2. The \\ enclosure regex used to check for strings ending in\\ enclosure  containing path separator\end{tabular} \\ \hline
\end{tabular}
\caption{Top 5 most frequently found CVE IDs from the un-identified security issues.}
\label{cve_5_description}
\end{table*}

\section{Details related to bots}\label{app_rq2}

An overview of various activities performed by bots in user-reported issues, along with their frequency in security-related issues, is provided in Table \ref{table:Frequency_of_botactivities}. Additionally, the classification of bots involved in user-reported security issues is detailed in Table \ref{tab:bot_classification_usercreated_issues}. A specific categorization of bot functionalities is further described in Table \ref{Bots_classes_description}. \snewtext{We also disclose a summary classification of 40 publicly available bots based on their implementation type. The Table \ref{bots-rulebased-cls} illustrates the range of bot types identified in our analysis.}

\begin{table}[]
\scriptsize
\centering
\begin{tabular}{|
>{\columncolor[HTML]{C0C0C0}}l |l|l|}
\hline
\textbf{Bots Type} & \cellcolor[HTML]{C0C0C0}\textbf{Description} & \cellcolor[HTML]{C0C0C0} \textbf{Example of bots (Functionality)} \\ \hline
 \textbf{Rule based (34 bots)} &  \begin{tabular}[c]{@{}l@{}}Implements deterministic \\ logic based on predefined \\ conditions and static rules.\end{tabular} &  \begin{tabular}[c]{@{}l@{}}mergify (PR management),  \\ google-cla (Policy \& CLA), \\ gatsby-cloud(CI/CD)\end{tabular} \\ \hline
 \textbf{AI/ML based (4 bots)} &  \begin{tabular}[c]{@{}l@{}}Uses learning-based \\ models to adapt behavior \\ from data rather than fixed \\ rules.\end{tabular} &  \begin{tabular}[c]{@{}l@{}}dotnetissuelabeler(issue labelling),\\ coderabbitai(pull request summarizer)\end{tabular} \\ \hline
 \textbf{In Beta stage (2 bots)} &  \begin{tabular}[c]{@{}l@{}}Includes experimental \\ AI/ML components \\ still under development\end{tabular} & \begin{tabular}[c]{@{}l@{}}sonarcloud (codefix, codereviews), \\ codecov (coverage of code)\end{tabular} \\ \hline
\end{tabular}
\caption{\snewtext{Classification of 40 publicly available bots based on implementation type, along with brief descriptions of their underlying logic.} }
\label{bots-rulebased-cls}
\end{table}

\begin{table*}[]
\scriptsize
\centering
\begin{tabular}{|l|l|l|}
\hline
\rowcolor[HTML]{C0C0C0} 
L1                                                               & L2                    & Description                                                                                                                          \\ \hline
\begin{tabular}[c]{@{}l@{}}Dependency \\ Management\end{tabular} &                       & \begin{tabular}[c]{@{}l@{}}Bots that automate the management and updates of project \\ dependencies.\end{tabular}                    \\ \hline
PR Management                                                    &                       & \begin{tabular}[c]{@{}l@{}}Bots that handle the entire pull request lifecycle, from creation \\ and review to merging and closure.\end{tabular}                       \\ \hline
CI/CD                                                            &                       & \begin{tabular}[c]{@{}l@{}}Bots that handle continuous integration and deployment \\ pipelines for codebases.\end{tabular}           \\ \hline
Security                                                         &                       & \begin{tabular}[c]{@{}l@{}}Bots focused on identifying vulnerabilities and ensuring \\ code security.\end{tabular}                   \\ \hline
                                                                 & Policy \& CLA          & \begin{tabular}[c]{@{}l@{}}Bots that enforce project policies or manage Contributor \\ License Agreements (CLA).\end{tabular}        \\ \cline{2-3} 
                                                                 & PR Related            & \begin{tabular}[c]{@{}l@{}}Bots that automate specific tasks within the pull request process,\\ such as labeling, commenting, or checking for conflicts.\end{tabular}                  \\ \cline{2-3} 
                                                                 & Project Management    & \begin{tabular}[c]{@{}l@{}}Bots that streamline task tracking, milestone planning, and \\ overall project coordination.\end{tabular} \\ \cline{2-3} 
                                                                 & Issue Related         & \begin{tabular}[c]{@{}l@{}}Bots that assist in tracking, labeling, or resolving issues \\ efficiently\end{tabular}                   \\ \cline{2-3} 
                                                                 & Code Coverage         & Bots that analyze and report on code test coverage metrics                                                                           \\ \cline{2-3} 
\multirow{-6}{*}{Utility}                                        & Miscellaneous Utility & \begin{tabular}[c]{@{}l@{}}Bots offering diverse functionalities that extend beyond \\ specific predefined categories.\end{tabular}  \\ \hline
Miscellaneous                                                    &                       & \begin{tabular}[c]{@{}l@{}}Bots that do not fit into the predefined categories and serve \\ diverse purposes\end{tabular}            \\ \hline
\end{tabular}
\caption{Description of each class of bot classification.}
\label{Bots_classes_description}
\end{table*}

\section{GLMM results} ~\label{app_RQ4}
We present three tables showcasing the results of the Generalized Linear Mixed Model (GLMM) analysis. These tables \ref{results_glmm_1}, \ref{results_glmm_2}, \ref{results_glmm_3} summarize the statistical outputs 

\begin{table}[]
\scriptsize
\centering

\begin{tabular}{|l|l|l|l|l|}
\hline
\rowcolor[HTML]{C0C0C0} 
\textbf{Factors}                                                        & \textbf{Category} & \textbf{O.R.} & \textbf{C.I.}      & \textbf{p-value}                    \\ \hline
Presence of CVE                                                         & -                 & 0.707         & {[}0.680, 0.735{]} & < $2 \times 10^{-16}$\\ \hline& 0-10              & 0.127         & {[}0.109, 0.147{]} & < $2 \times 10^{-16}$ \\ \cline{2-5}  & 10-25             & 0.363         & {[}0.312, 0.423{]} & < $2 \times 10^{-16}$ \\ \cline{2-5} & 25-50             & 0.568         & {[}0.486, 0.664{]} &  $ 1.18\times 10^{-12}$\\ \cline{2-5} 
\multirow{-4}{*}{Number of Comments}                                    & 50-80             & 0.748         & {[}0.624, 0.898{]}  & $ 1.82\times 10^{-3}$         \\ \hline
\begin{tabular}[c]{@{}l@{}}Reproducibility of \\ the issue\end{tabular} & -                 & 2.213         & {[}2.196, 2.231{]} & < $2 \times 10^{-16}$ \\ \hline
\end{tabular}
\caption{Results of Generalized Linear Mixed Model(glmm) examining the factors for time to close security related issues.}
\label{results_glmm_1}
\end{table}

\begin{table}[]
\scriptsize
\centering
\begin{tabular}{|l|l|l|l|l|}
\hline
\rowcolor[HTML]{C0C0C0} 
\textbf{Factors} & \textbf{Category} & \textbf{O.R.} & \textbf{C.I.} & \textbf{p-value} \\ \hline
Presence of CVE & - & 0.775 & {[}0.688, 0.873{]} & $2.83 \times 10^{-5}$ \\ \hline
 & 0-10 & 2.256 & {[}1.307, 3.892{]} & $3.472 \times 10^{-3}$ \\ \cline{2-5} 
 & 10-25 & 2.561 & {[}1.482, 4.425{]} & $7.48 \times 10^{-4}$ \\ \cline{2-5} 
\multirow{-3}{*}{Number of Comments} & 25-50 & 2.088 & {[}1.197, 3.643{]} & $9.51 \times 10^{-3}$ \\ \hline
\begin{tabular}[c]{@{}l@{}}Number of weekly\\ downloads\end{tabular} & 50K-100K & 1.619 & {[}1.209, 2.168{]} & $1.216 \times 10^{-3}$ \\ \hline
\begin{tabular}[c]{@{}l@{}}Reproducibility of \\ the issue\end{tabular} & - & 5.060 & {[}4.931, 5.191{]} & < $2 \times 10^{-16}$ \\ \hline
Involvement of Bots & - & 14.602 & {[}14.192, 15.023{]} & < $2 \times 10^{-16}$ \\ \hline
\end{tabular}
\caption{Results of Generalized Linear Mixed Model(glmm) examining the factors for staleness of security related issues.}
\label{results_glmm_2}
\end{table}

\begin{table}[]
\scriptsize
\centering
\begin{tabular}{|l|l|l|l|l|}
\hline
\rowcolor[HTML]{C0C0C0} 
\textbf{Factors} & \textbf{Category} & \textbf{O.R.} & \textbf{C.I.} & \textbf{p-value} \\ \hline
Presence of CVE & - & 0.816 & {[}0.786, 0.848{]} & < $2 \times 10^{-16}$ \\ \hline
Number of Comments & 10-25 & 0.781 & {[}0.669, 0.912{]} &  $1.74 \times 10^{-3}$ \\ \hline
 & 0K-10K & 1.267 & {[}1.225, 1.311{]} & < $2 \times 10^{-16}$ \\ \cline{2-5} 
\multirow{-2}{*}{\begin{tabular}[c]{@{}l@{}}Number of weekly\\ downloads\end{tabular}} & 10K-50K & 1.073 & {[}1.027, 1.121{]} &  $1.65 \times 10^{-3}$ \\ \hline
\begin{tabular}[c]{@{}l@{}}Reproducibility of \\ the issue\end{tabular} & - & 0.419 & {[}0.415, 0.423{]} & < $2 \times 10^{-16}$ \\ \hline
Involvement of Bots & - & 0.682 & {[}0.677, 0.687{]} & < $2 \times 10^{-16}$ \\ \hline
\end{tabular}
\caption{Results of Generalized Linear Mixed Model(glmm) examining the factors for successful resolution of security related issues.}
\label{results_glmm_3}
\end{table}

\begin{table}[h]
\scriptsize
\centering
\begin{tabular}{|p{2.5cm}|p{1cm}|p{4.5cm}|}
\hline
\rowcolor[HTML]{C0C0C0} 
\textbf{Factor considered}                 & \textbf{Type} & \cellcolor[HTML]{C0C0C0}\textbf{Description}                                                                                                                     \\ \hline
CVE mention                                & Boolean       & Refers to mention of CVE/CWE in the issue                                                                                                                        \\ \hline
Number of comments                         & Categorical   & Number of comments on the issue categorised into 5 classes: i) 0-10, ii) 10-25, iii) 25-50, iv) 50-80, v) \textgreater{}80                                       \\ \hline
Weekly downloads                           & Categorical   & No. of weekly downloads of the package categorised into 5 classes: i) 0-10K, ii) 10K-50K, iii) 50K-100K, iv) 100K-200K, v) \textgreater{}200K                    \\ \hline
No. of active maintainers in the past year & Categorical   & No. of active maintainers of the repository in the last one year, categorised into 5 classes: i) 0-10, ii) 10-50, iii) 50-100, iv) 100-200, v) \textgreater{}200 \\ \hline
Reproducibility                            & Boolean       & If the issue body has explicit instructions on reproducing the error like code snippets, step-by-step instructions or error logs etc.                            \\ \hline
Bot Involvement                            & Boolean       & If bot is involved in the discussion or closing of the issue                                                                                                     \\ \hline
\end{tabular}
\caption{Description of factors considered in GLMM models.}
\label{tab:factor_desc}
\end{table}

\end{document}